\documentclass[11pt]{article}
\usepackage{hyperref}
\usepackage{tikz}
\usepackage{multicol}
\usepackage{mdframed}
\usepackage{color}
\usepackage{epsfig,subfigure}
\usepackage{kbordermatrix}
\usepackage{amssymb, amsmath}
\usepackage{color}
\usepackage{bm}
\usepackage{amsthm}
\usepackage{authblk}
\usepackage{graphicx}
\usepackage{color}

\usepackage{amssymb}
\usepackage{amsmath,amsfonts,amssymb}
\usepackage{verbatim}
\usepackage{stfloats}
\usepackage[bookmarks=false]{}

\newtheorem{theorem}{Theorem}
\newtheorem{corollary}{Corollary}
\newtheorem{definition}{Definition}
\newtheorem{lemma}{Lemma}

\begin{document}

\title{On the Asymptotic Capacity of $X$-Secure $T$-Private\\ Information Retrieval with  Graph Based Replicated Storage}
\author{Zhuqing Jia and Syed A. Jafar}
\affil{Center for Pervasive Communications and Computing (CPCC), UC Irvine\\
Email: \{zhuqingj, syed\}@uci.edu}
\date{}
\maketitle

\begin{abstract} 
The problem of private information retrieval with graph-based replicated storage was recently introduced by Raviv, Tamo and Yaakobi. Its capacity remains open in almost all cases. In this work  the asymptotic  (large number of messages) capacity of this problem  is studied along with its generalizations to include arbitrary $T$-privacy and $X$-security constraints, where the privacy of the user must be protected against any set of up to $T$ colluding servers and the security of the stored data must be protected against any set of up to $X$ colluding servers. A general achievable scheme for arbitrary storage patterns is presented that achieves the rate $(\rho_{\min}-X-T)/N$, where $N$ is the total number of servers, and each message is replicated at least $\rho_{\min}$ times. Notably, the scheme makes use of a special structure inspired by dual Generalized Reed Solomon  (GRS) codes. A general converse is also presented. The two bounds are shown to match for many settings, including  symmetric storage patterns. Finally, the asymptotic capacity is fully characterized for the case without security constraints $(X=0)$ for arbitrary storage patterns provided that each message is replicated no more than $T+2$ times. As an example of this result, consider PIR with arbitrary graph based storage ($T=1, X=0$) where every message is replicated at exactly $3$ servers. For this $3$-replicated storage setting, the asymptotic capacity is  equal to $2/\nu_2(G)$ where $\nu_2(G)$ is the maximum size of a $2$-matching in a storage graph $G[V,E]$. In this undirected graph, the vertices $V$ correspond to the set of servers, and there is an edge $uv\in E$ between vertices $u,v$ only if a subset of messages is replicated at both servers $u$ and $v$. 
\end{abstract}

\newpage
\section{Introduction}
As distributed storage systems become increasingly prevalent, there are mounting concerns regarding user privacy and data security. The problem of $X$-secure and $T$-private information retrieval (XSTPIR) deals with both of these issues \cite{Jia_Sun_Jafar_XSTPIR}.  In its basic form, private information retrieval (PIR) involves $K$ datasets (messages) that are replicated at $N$ distributed servers, and a user who wishes to retrieve one of these datasets without revealing any information about the identity of his desired dataset to any of the servers \cite{PIRfirst, PIRfirstjournal}. XSTPIR is a generalization of PIR where the stored data must remain secure as long as the number of colluding servers is not more than $X$, and the user's privacy must be preserved as long as the number of colluding servers is not more than $T$ \cite{Jia_Sun_Jafar_XSTPIR}. The rate of a PIR scheme is the ratio of the number of bits of desired message that are  retrieved per bit of total download from all servers. The supremum of achievable rates is called the capacity of PIR \cite{Sun_Jafar_PIR}.

The capacity of the basic PIR setting was  characterized in \cite{Sun_Jafar_PIR} for arbitrary number of messages replicated across arbitrary number of servers. Following in the footsteps of \cite{Sun_Jafar_PIR} there has been a wave of new results exploring the fundamental limits of PIR under a variety of constraints. This includes PIR with $T$-privacy and replicated storage \cite{Sun_Jafar_TPIR},  PIR with MDS coded storage \cite{Tajeddine_Rouayheb,Banawan_Ulukus}, PIR with optimal storage and upload cost \cite{Tian_Sun_Chen_Upload}, PIR with arbitrary message lengths \cite{Sun_Jafar_PIRL},  PIR with restricted collusion patterns \cite{Tajeddine_Gnilke_Karpuk_Etal, Jia_Sun_Jafar}, PIR with $T$-privacy and MDS coded storage \cite{FREIJ_HOLLANTI, Sun_Jafar_MDSTPIR}, multi-message PIR \cite{Banawan_Ulukus_Multimessage}, PIR with asymmetric traffic constraints \cite{Banawan_Ulukus_Traffic}, multi-round PIR \cite{Sun_Jafar_MPIR}, cache-aided and otherwise storage-constrained PIR \cite{Tandon_CPIR, Wei_Banawan_Ulukus}, PIR with side-information \cite{Kadhe_Garcia_Heidarzadeh_Rouayheb_Sprintson, Chen_Wang_Jafar}, PIR for computation \cite{Sun_Jafar_PC, Mirmohseni_Maddah, obead2018achievable, David_Karpuk}, PIR for security against eavesdroppers \cite{Banawan_Ulukus_Asymmetric, Wang_Sun_Skoglund}, PIR with Byzantine adversaries \cite{Banawan_Ulukus_Byzantine, Zhang_Ge_Variant, Tajeddine_Gnilke_Karpuk_Hollanti}, symmetrically secure PIR \cite{Sun_Jafar_SPIR, Wang_Skoglund_TSPIR, Wang_Skoglund_SPIRAd}, and   PIR with secure storage \cite{Yang_Shin_Lee, Jia_Sun_Jafar_XSTPIR}.

 Most relevant to this work is the recent characterization in \cite{Jia_Sun_Jafar_XSTPIR} of the asymptotic ($K\rightarrow\infty$) capacity of XSTPIR as $C_{\mbox{\tiny XSTPIR}}=1-(X+T)/N$.  Note that the XSTPIR setting includes as special case the TPIR setting, obtained by setting $X=0$, as well as the original PIR setting, obtained by setting $X=0$ and $T=1$. It is limited, however, by its assumption of fully replicated storage, i.e.,  all messages are stored by all servers, which can be burdensome for  large data sets. Motivated by the preference for simple storage, Raviv, Tamo and Yaakobi in \cite{Raviv_Tamo_Yaakobi} introduced a graph based replicated storage model. Instead of full replication where every message is replicated at every server, graph based replication assumes that each message is replicated only among a subset of servers. This allows a graph representation where the  vertices are the $N$ servers and each message is represented by a hyperedge comprised of vertices (servers)  where this message is replicated. Reference \cite{Raviv_Tamo_Yaakobi}  primarily focuses on GTPIR, i.e., PIR with graph based replicated storage and $T$-privacy. An achievable scheme is proposed that achieves the rate $1/N$ as long as $T$ is smaller than the replication factor of each message (the number of servers where the message is replicated), and is shown to be within a factor of $2$ from optimality for some special cases. However, optimal GTPIR schemes remain unknown in almost all settings. Understanding the key ideas that constitute optimal PIR schemes under graph based replicated storage is our goal  in this paper.
 
 The main contributions of this work are as follows. We study the asymptotic capacity of $T$-private and $X$-secure PIR with graph-based replicated storage, in short GXSTPIR. Recall that asymptotic capacity is quite meaningful for PIR because the number of messages is typically large, and the convergence of capacity to its asymptotic value tends to take place quite rapidly \cite{Jia_Sun_Jafar_XSTPIR}. GXSTPIR includes as special cases the settings of GTPIR \cite{Raviv_Tamo_Yaakobi}, XSTPIR \cite{Jia_Sun_Jafar_XSTPIR}, TPIR \cite{Sun_Jafar_TPIR} and basic PIR \cite{Sun_Jafar_PIR}, and as such it presents a unified view of these settings. Our first result is an achievable scheme for GXSTPIR  that achieves the rate $(\rho_{\min}-X-T)/N$ for arbitrary storage patterns provided every message is replicated at least $\rho_{\min}$ times. In addition to ideas like cross-subspace alignment, Reed-Solomon (RS) coded storage and RS coded queries that were previously used for XSTPIR \cite{Jia_Sun_Jafar_XSTPIR}, a key novelty of our achievable scheme for GXSTPIR is how it creates and takes advantage of a structure inspired by dual Generalized Reed Solomon (GRS) codes. This is explained intuitively in Section \ref{sec:grsintuit}. Our second contribution is a general converse bound for asymptotic capacity of GXSTPIR with arbitrary storage patterns. While the asymptotic capacity of GXSTPIR remains open in general, it is remarkable that our converse bound is tight in all settings where we are able to settle the capacity. In particular, the general achievable scheme matches the converse bound when the storage is symmetric, settling the asymptotic capacity for those settings. For several examples with asymmetric storage, it turns out that the achievable scheme can be improved to match the converse bound by applying it only after eliminating certain redundant servers. Thus, the asymptotic capacity for such cases is settled as well. In general however, with arbitrary graph based storage, more sophisticated achievable schemes may be obtained by combining our achievable scheme with ideas from private computation \cite{Sun_Jafar_PC}. To illustrate this, we consider the GTPIR problem ($X=0$) where every message is replicated no more than $T+2$ times. As our final result, for this problem we fully settle the asymptotic capacity for arbitrary storage patterns. The asymptotic capacity depends strongly on the storage graph structure, and requires a private computation scheme on top of our general achievable scheme. As an example of this result, consider GPIR, i.e., PIR with arbitrary graph based storage ($T=1, X=0$) where every message is replicated at exactly $3$ servers. For this $3$-replicated storage setting, the asymptotic capacity is exactly equal to $2/\nu_2(G)$ where $\nu_2(G)$ is the maximum size of a $2$-matching in a storage graph $G[V,E]$. In this storage graph, the vertices $V$ correspond to the set of servers, and there is an edge $uv\in E$ between vertices $u,v$ only if a subset of messages is replicated at both servers $u$ and $v$. . This is consistent with the intuition that storage graph properties must be essential to the asymptotic capacity of graph-based storage.

\emph{Notation: }For a positive integer $M$ the notation $[M]$ denotes the set $\{1,2,\cdots,M\}$. The notation $X_{[M]}$ stands for the set  $\{X_1,X_{2},\dots,X_M\}$. Similarly, for an index set $\mathcal{I}=\{i_1,i_2,\dots,i_n\}$, $X_{\mathcal{I}}$ denotes the set $\{X_{i_1},X_{i_2},\dots,X_{i_n}\}$. If $A$ is a set of random variables, then by $H(A)$ we denote the joint entropy of those random variables. Mutual informations between sets of random variables are similarly defined. For tuples such as $A=(a_1, a_2, \cdots, a_n)$ we allow set theoretic notions of inclusion. For example, $b\in A$ denotes the relationship $b\in\{a_1, a_2, \cdots, a_n\}$. Similarly, $b\in A\setminus \{a_1\}$ denotes $b\in\{a_2,a_3, \cdots, a_n\}$. The notation $X\sim Y$ is used to indicate that $X$ and $Y$ are identically distributed. When a natural number, say $\ell\in\mathbb{N}$, is used to represent an element of a finite field $\mathbb{F}_q$, it denotes the sum of $\ell$ ones in $\mathbb{F}_q$, i.e., $\ell\triangleq \sum_{l=1}^\ell 1$, where the addition is over $\mathbb{F}_q$.

\section{Problem Statement}
We begin with a description of messages and storage structure. Based on the storage structure we will partition the set of messages into $M$ subsets so that the messages in the same subset have the same storage structure. Define $\mathcal{W}=(\mathcal{W}_1,\mathcal{W}_2, \cdots, \mathcal{W}_M)$ where $\mathcal{W}_m, m\in[M]$, is comprised of $K_m$ messages,
\begin{align}
\mathcal{W}_m&=(W_{m,1}, W_{m,2},\cdots, W_{m,K_m}).
\end{align}
Messages are independent, and each message is composed of $L$ i.i.d. uniform symbols from $\mathbb{F}_q$, i.e.,
\begin{align}
&H(W_{m,k})=H(W_{m,k}(1), W_{m,k}(2), \cdots, W_{m,k}(L))=L,& ~\forall m\in[M], k\in[K_m]\\
&H(W_{1,1}, \cdots, W_{M,K_M})=\sum_{m=1}^{M}K_mL,
\end{align}
in $q$-ary units. There are a total of $N$ servers. Corresponding to $\mathcal{W}=(\mathcal{W}_1,\cdots, \mathcal{W}_M)$, let us define 
\begin{align}
\mathcal{R}&=(\mathcal{R}_1, \cdots, \mathcal{R}_M),\\
\mathcal{R}_m&=\left(\mathcal{R}_m(1), \cdots, \mathcal{R}_m(\rho_m)\right), \forall m\in[M],\\
\mathcal{R}_m(r)&\in[N], \forall r\in[\rho_m],
\end{align}
where $\mathcal{R}_m, m\in[M]$ contains the servers, $\mathcal{R}_m(r)\in[N]$ that store the $m^{th}$ set of messages $\mathcal{W}_m$.  Without loss of generality we will assume that the servers are listed in increasing order in each tuple $\mathcal{R}_m$. The cardinality of $\mathcal{R}_m$ is $|\mathcal{R}_m|=\rho_m$, which will be referred to as the replication factor for the messages in $\mathcal{W}_m$. The minimum replication factor is defined as
\begin{align}
\rho_{\min}&\triangleq \min_{m\in[M]}\rho_m.
\end{align}
It is important to note that  the messages may not be directly replicated at the servers. Because of security constraints, each message ${W}_{m,k}\in \mathcal{W}_m$, is represented by a total of $\rho_m$ \emph{shares} (the nomenclature comes from secret-sharing), denoted $\overline{W}_{m,k}=\left({W}_{m,k}^{(n)}, n\in\mathcal{R}_m\right)$, such that the share ${W}_{m,k}^{(n)}$ is stored at Server $n$,  for all  $n\in\mathcal{R}_m$. Messages are independently secured and must be recoverable from their shares, as specified by the following constraints.
\begin{align}
H\left(\overline{W}_{1,1}, \cdots, \overline{W}_{M,K_m}\right)&=\sum_{m\in[M], k\in[K_M]}H\left(\overline{W}_{m,k}\right),\label{shareindp}\\
H\left(W_{m,k}\mid \overline{W}_{m,k}\right)&=0.\label{eq:wfromshare}
\end{align}
The information stored at Server $n$ is defined as
\begin{align}
S_n&=\left\{W_{m,k}^{(n)}, m\in[M], k\in[K_m], \mathcal{R}_m\ni n\right\}.\label{eq:stor}
\end{align}
Let us also define the index set of $\mathcal{W}_m$ that are stored at Server $n$, as
\begin{align}
\mathcal{M}_n&=\{m\in[M] \Big| \mathcal{R}_m\ni n\}.
\end{align}

\noindent For example, suppose we have $M=4$ message sets  (each comprised of $K_m=2$ messages), stored at $N=4$ servers as shown.

\bigskip

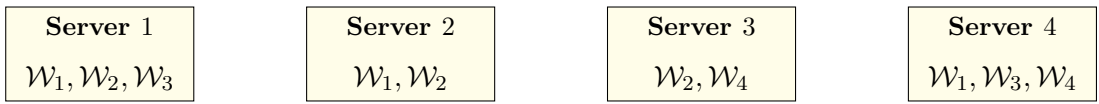
\begin{center}
\begin{tikzpicture}
\node (server1) [rectangle, fill=yellow!10, draw, minimum width=2.5cm, minimum height=1.25cm, anchor= south west] at (0,0) {};
\node[anchor=south] at (server1.south) {$\mathcal{W}_1,\mathcal{W}_2,\mathcal{W}_3$};
\node[anchor=north] at (server1.north) {\small \bf Server $1$};
\node (server2) [rectangle, fill=yellow!10, draw, minimum width=2.5cm, minimum height=1.25cm, anchor= south west] at (4,0) {};
\node[anchor=south] at (server2.south) {$\mathcal{W}_1,\mathcal{W}_2$};
\node[anchor=north] at (server2.north) {\small \bf Server $2$};
\node (server3) [rectangle, fill=yellow!10, draw, minimum width=2.5cm, minimum height=1.25cm, anchor= south west] at (8,0) {};
\node[anchor=south] at (server3.south) {$\mathcal{W}_2,\mathcal{W}_4$};
\node[anchor=north] at (server3.north) {\small \bf Server $3$};
\node (server4) [rectangle, fill=yellow!10, draw, minimum width=2.5cm, minimum height=1.25cm, anchor= south west] at (12,0) {};
\node[anchor=south] at (server4.south) {$\mathcal{W}_1,\mathcal{W}_3,\mathcal{W}_4$};
\node[anchor=north] at (server4.north) {\small \bf Server $4$};
\end{tikzpicture}
\end{center}

\noindent Then for this example,\footnote{Incidentally, our results will show that as $K_m\rightarrow \infty$,  for this example  $C_\infty=1/3$, and Server $2$ is redundant.} we have,
\begin{align}
\mathcal{M}_1&=\{1,2,3\},&S_1&=\{W_{1,1}^{(1)}, W_{1,2}^{(1)},W_{2,1}^{(1)}, W_{2,2}^{(1)},W_{3,1}^{(1)}, W_{3,2}^{(1)}\},
&\mathcal{R}_1&=(1,2,4),&\rho_1=3,\label{eq:syma}\\
\mathcal{M}_2&=\{1,2\},&S_2&=\{W_{1,1}^{(2)}, W_{1,2}^{(2)},W_{2,1}^{(2)}, W_{2,2}^{(2)}\},&\mathcal{R}_2&=(1,2,3), &\rho_2=3,\\
\mathcal{M}_3&=\{2,4\},&S_3&=\{W_{2,1}^{(3)}, W_{2,2}^{(3)},W_{4,1}^{(3)}, W_{4,2}^{(3)}\},&\mathcal{R}_3&=(1,4), &\rho_3=2,\\
\mathcal{M}_4&=\{1,3,4\},&S_4&=\{W_{1,1}^{(4)}, W_{1,2}^{(4)},W_{3,1}^{(4)}, W_{3,2}^{(4)},W_{4,1}^{(4)}, W_{4,2}^{(4)}\},&\mathcal{R}_4&=(3,4), &\rho_4=2,\label{eq:symz}
\end{align}
and $\rho_{\min}=2$.

 The $X$-secure constraint, $0\leq X\leq N$, requires that any $X$ (or fewer) colluding servers learn nothing about the messages.
\begin{align}
\text{[$X$-Security] }&& I(S_{\mathcal{X}};\mathcal{W})&=0, &&\forall \mathcal{X}\subset [N], |\mathcal{X}|\leq X. \label{eq:secur}
\end{align}
$X=0$ represents the setting without security constraints. If $X=0$, then no secret sharing is needed, so each share of a message is the message itself, 
\begin{align}
X=0&\implies W_{m,k}^{(n)}=W_{m,k}, &&\forall n\in\mathcal{R}_m.
\end{align}
 This completes the description of the messages and  the storage at the $N$ servers. Next, let us describe the private information retrieval aspect.

The user desires the message $W_{\mu,\kappa}$, where the indices $\mu$ and $\kappa$ are chosen privately and uniformly by the user from $\mu\in[M],\kappa\in[K_\mu]$, respectively. In order to retrieve his desired message, the user generates $N$ queries, $Q_1^{[\mu,\kappa]},Q_2^{[\mu,\kappa]},\dots,Q_N^{[\mu,\kappa]}$, and sends the $n^{th}$ query, $Q_n^{[\mu,\kappa]}$ to the $n$-th server. The user has no prior knowledge of the message realizations,
\begin{equation}
I\left(S_{[N]}~;~\mu,\kappa, Q_{[N]}^{[1,1]}, \cdots, Q_{[N]}^{[M,K_M]}\right)=0.\label{eq:indp}
\end{equation}
A $T$-private scheme, $1\leq T\leq N$, requires that any $T$ (or fewer) colluding servers learn nothing about $(\mu,\kappa)$.
\begin{align}
\text{[$T$-Privacy] }&&I\left(Q_{\mathcal{T}}^{[\mu,\kappa ]}~; ~\mu,\kappa\right)&=0, &&\forall \mathcal{T}\subset [N], |\mathcal{T}|\leq T.\label{eq:tpriv}
\end{align}
Upon receiving the query $Q_n^{[\mu,\kappa]}$, the $n$-th server generates an answer string $A_n^{[\mu,\kappa]}$, which is a function of the query $Q_n^{[\mu,\kappa]}$ and its stored information $S_n$. 
\begin{align}
H\left(A_n^{[m,k]}\mid Q_n^{[m,k]},S_n\right)&=0, &&\forall m\in[M], k\in[K_m].\label{eq:ansfunc}
\end{align}
The correctness constraint guarantees that from all the answers, the user is able to decode the desired message $W_{\mu,\kappa}$,
\begin{align}
\text{[Correctness] }&& H\left(W_{\mu,\kappa} \mid A_{[N]}^{[\mu,\kappa]},Q_{[N]}^{[\mu,\kappa]},\mu,\kappa\right)=0. \label{corr}
\end{align}
The rate of a GXSTPIR  scheme is defined by the number of $q$-ary symbols of desired message that are retrieved per downloaded $q$-ary symbol,
\begin{equation}
R= \frac{H(W_{\mu,\kappa})}{\sum_{n\in[N]}H\left(A_n^{[\mu,\kappa]}\right)}=\frac{L}{D},
\end{equation}
where $D=\sum_{n\in[N]}H\left(A_n^{[\mu,\kappa]}\right)$ is the expected total number of $q$-ary symbols downloaded by the user from all servers.
The capacity of GXSTPIR, denoted as $C(N, X, T, \mathcal{W}, \mathcal{S})$, is the supremum of $R$ across all feasible schemes. In this work we are interested in the setting where each subset of messages is comprised of a large number of messages. Specifically, we wish to characterize the asymptotic capacity, as $K_m\rightarrow\infty$ for all $m\in[M]$. In order to have $K_m$ approach infinity together for all $m\in[M]$, let us define,
\begin{align}
K_{\min}&=\lceil \chi_m K \rceil,
\end{align}
so that $\chi_m, m\in[M]$ are fixed  constants, while $K$ approaches infinity.
Then the asymptotic capacity is defined as
\begin{align}
C_\infty=\lim_{K\rightarrow\infty}C(N,X,T,\mathcal{W},\mathcal{S}).
\end{align}
Note that the number of message sets, $M$, and the storage pattern $\mathcal{R}$ remain unchanged, while $K_m$, i.e., the number of messages in each $\mathcal{W}_m$  approaches infinity.

\allowdisplaybreaks
\section{Results}
Our first result is a general achievability argument that provides us a lower bound on the asymptotic capacity of GXSTPIR. 
\begin{theorem}\label{thm:ach}
The asymptotic capacity of GXSTPIR is bounded below as follows,
\begin{align}
C_\infty&\geq\frac{\rho_{\min}-X-T}{N}.
\end{align}
\end{theorem}
The proof of Theorem \ref{thm:ach} appears in Section \ref{proof:ach}. An interesting aspect of the proof is the use of a structure inspired by dual GRS codes, that is intuitively explained in Section \ref{sec:grsintuit}. Another interesting aspect of Theorem \ref{thm:ach} is that  applying it to a subset of servers (by eliminating the rest) may produce a higher achievable rate than if all servers were used. Therefore, in order to find the best achievable rate guaranteed by Theorem \ref{thm:ach} we must choose the best subset of servers. Example \ref{ex:redserv} in Section \ref{sec:examples} illustrates this idea.

 Our next result is a converse argument that  holds for arbitrary storage patterns. Recall that $D_n=H(A_n^{[\mu,\kappa]})/L$ is the normalized download from Server $n$.  
\begin{theorem}\label{thm:converse}
The asymptotic capacity of GXSTPIR is bounded above as follows,
\begin{align}
C_\infty&\leq \left\{
\begin{array}{ll}
0,&\rho_{\min}\leq X+T\\
\max_{(D_1, \cdots, D_N)\in\mathcal{D}}~~~ \left(D_1+D_2+\cdots+D_N\right)^{-1}, &\rho_{\min}> X+T
\end{array}
\right.
\\
\intertext{and $\mathcal{D}$ is defined as}
\mathcal{D}&\triangleq\left\{(D_1, \cdots, D_N)\in\mathbb{R}_+^N~\Big|~ \sum_{n\in \mathcal{R}_m'} D_n\geq 1, ~\forall m\in[M],  \mathcal{R}_m'\subset\mathcal{R}_m, |\mathcal{R}_m'|=|\mathcal{R}_m|-X-T \right\}\label{eq:defD}.
\end{align}
\end{theorem}
The proof of Theorem \ref{thm:converse} appears in Section \ref{proof:converse}.  Since the asymptotic capacity is zero for $\rho_{\min}\leq X+T$, in the remainder of this section we will assume that $\rho_{\min}>X+T$.

{\it Remark:} Note that \eqref{eq:defD} implies that the total normalized download from any  $\rho_m-X-T$ servers in $\mathcal{R}_m$ must be at least $1$. A simple averaging argument implies that the total normalized download from all $\rho_m$ servers in any $\mathcal{R}_m$ must be at least $\rho_m/(\rho_m-X-T)$.

The general lower bound in Theorem \ref{thm:ach} is in closed form and the general upper bound in Theorem \ref{thm:converse} is essentially a linear program, so for arbitrary settings it is possible to evaluate both  to check if they match (provided the parameter values are not too large to be computationally feasible). Conceptually, the condition for them to match may be understood as follows. Consider a hypergraph $\mathcal{G}(\mathcal{V},\mathcal{E})$ with the set of vertices $\mathcal{V}=[N]$ representing the $N$ servers, and the set of hyperedges $\mathcal{E}$ such that $e\in\mathcal{E}$ if and only if $\exists m\in[M]$ such that $e\subset\mathcal{R}_m$ and $|\mathcal{R}_m|-|e|=X+T$. For this graph, hyperedges $e\in\mathcal{E}$,  with corresponding weights $x_e\in\mathbb{R}_+$,  are said to form a fractional matching if for every vertex $v\in\mathcal{V}$ the total weight of the edges that include $v$ is less than or equal to $1$. The largest possible total weight of a fractional matching is called the fractional matching number of $\mathcal{G}$ \cite{Schrijver}. As shown in Lemma \ref{lemma:frac} in Appendix \ref{app:lemmas}, the optimal converse bound from Theorem \ref{thm:converse} on the total normalized download, i.e.,  $\min_{\mathcal{D}}(D_1+\cdots+D_N)$ is equal to the fractional matching number of $\mathcal{G}[\mathcal{V},\mathcal{E}]$. Thus, the following corollary immediately follows.

\begin{corollary} \label{cor:frac} The lower bound of Theorem \ref{thm:ach} matches the upper bound of Theorem \ref{thm:converse} if and only if the fractional matching number of $\mathcal{G}(\mathcal{V},\mathcal{E})$ is equal to $\frac{N}{\rho_{\min}-X-T}$. For all such cases, the asymptotic capacity $C_\infty=(\rho_{\min}-X-T)/N$.
\end{corollary}

Next let us identify some interesting special cases of Corollary \ref{cor:frac}.

Let $\mathcal{R}_{\mathcal{M}'}$ be a collection of the sets $\mathcal{R}_m, m\in \mathcal{M}'\subset[M]$.
We define $\mathcal{R}_{\mathcal{M}'}$ to be an exact $b$-cover of $[N]$ if $\rho_m=\rho_{\min}$ for all $m\in \mathcal{M}'$, and every element of $[N]$ is contained in exactly $b$ sets in $\mathcal{R}_{\mathcal{M}'}$. It follows that the asymptotic capacity $C_\infty=(\rho_{\min}-X-T)/N$ if there exists an exact $b$-cover for some $b\in\mathbb{Z}_+$. This is easily seen because for each $\mathcal{R}_m$ in $\mathcal{R}_{\mathcal{M}'}$ we have the bound $\sum_{n\in\mathcal{R}_m}D_n\geq \rho_{\min}/(\rho_{\min}-X-T)$ according to \eqref{eq:defD}. Adding all these bounds we obtain the desired converse bound $b\sum_{n\in[N]}D_n\geq (bN/\rho_{\min})(\rho_{\min}/(\rho_{\min}-X-T))$, i.e., $\sum_{n\in[N]}D_n\geq N/(\rho_{\min}-X-T)$, which is  achievable according to Theorem \ref{thm:ach}. 

As a special case that is of particular interest, define a symmetric storage setting as one where (after some permutation of message and server indices) for all $m\in[M]$, $\mathcal{R}_m=(\rho m+1, \rho m+2, \cdots, \rho m+\rho_{\min})$. Here, $\rho\leq \rho_{\min}$ and server indices are interpreted modulo $N$, e.g., Server $N+1$ is the same as Server $1$. Furthermore, $b=M\rho_{\min}/N$ is an integer value. Then any symmetric storage setting thus defined  has asymptotic capacity  $C_\infty=(\rho_{\min}-X-T)/N$ because the storage sets form an exact $b$-cover. 

Based on these observations, here are some examples of storage patterns where the asymptotic capacity is $C_\infty=(\rho_{\min}-X-T)/N$.

\begin{enumerate}
\item $\mathcal{R}=((1,2),(2,3),(3,1))$ which is a symmetric storage setting (forms an exact $2$ cover).
\item $\mathcal{R}=((1,2,3),(3,4,5),(5,1,2),(2,3,4),(4,5,1))$ which is a symmetric storage setting (forms an exact $3$-cover).
\item $\mathcal{R}=((1,2),(2,3),(3,1),(4,5),(5,6),(6,4))$ because it forms an exact $2$ cover.
\item $\mathcal{R}=((1,2,3),(4,5,6),(i,j,k),(a,b,c,d))$ for arbitrary $\{i,j,k\},\{a,b,c,d\}\subset[N]=[6]$ because it contains an exact $1$-cover, $\mathcal{R}_{\mathcal{M}'}=\{(1,2,3),(4,5,6)\}$.
\item $\mathcal{R}=((1,2,3),(3,4,1),(2,5,6),(4,5,6),(1,3,6),(1,2,5,6))$ because it contains an exact $2$-cover of $[N]=[6]$ in $\mathcal{R}_{\mathcal{M}'}=\{(1,2,3),(3,4,1),(2,5,6),(4,5,6)\}$.
\end{enumerate}
While the existence of an exact $b$-cover for some positive integer $b$ is \emph{sufficient} to guarantee that the asymptotic capacity is $C_\infty=(\rho_{\min}-X-T)/N$, it is not a \emph{necessary} condition. Examples \ref{ex:symnocover1} and \ref{ex:symnocover2} in Section \ref{sec:examples} show such settings.

On the other hand, it is also easy to see  that the lower bound of Theorem \ref{thm:ach} and the upper bound of Theorem \ref{thm:converse} do not always match.
Remarkably, in all such cases that we have been able to settle so far, it is the upper bound that is tight, and the achievability that needs to be improved. In many cases, such as Example \ref{ex:redserv} in Section \ref{sec:examples}, an improved achievability result is found easily by eliminating a redundant server before applying Theorem \ref{thm:ach}. However, more sophisticated achievable schemes may be required in general.

Our final result emphasizes this point by settling the asymptotic capacity of GTPIR, i.e., $T$-private information retrieval with arbitrary graph based storage and no security constraints $(X=0)$, provided each message is replicated no more than $(T+2)$ times. Because this result deals with arbitrary storage patterns, for its precise statement we will need the following definitions that follow the convention of  Schrijver \cite{Schrijver}.

\begin{definition}
Define $G=(V, E)$ as a simple undirected graph with vertices $V=[N]$  corresponding to the $N$ servers, and with edges $uv\in E$ if and only if $\{u,v\}\subset \mathcal{R}_m$ for some $m\in[M]$.
\end{definition}

\begin{definition} A set $U\subset V$ is called a stable set (also called independent set) if there are no edges between any two members of $U$.
\end{definition}

\begin{definition}
For $U\subset [N]$, define $\mathcal{N}(U)$ as the set of vertices in $V\backslash U$ that are neighbors of vertices in $U$. 
\end{definition}

\begin{definition}
Define $\delta(n)$ as the set of edges incident with vertex $n$.
\end{definition}

\begin{definition}
A function $x:E\rightarrow \mathbb{Z}_+$ is denoted as a vector $x\in \mathbb{Z}_+^E$. A function $y:V\rightarrow \mathbb{Z}_+$ is similarly denoted as a vector $y\in \mathbb{Z}_+^V$. The size of a vector is defined as the sum of its entries.
\end{definition}

\begin{definition}
For any $x\in\mathbb{Z}_+^E$, and $F\subset E$, define $x(F)=\sum_{f\in F}x(f)$. 
\end{definition}

\begin{definition}
A $b$-matching in $G$ is defined as a vector $x\in\mathbb{Z}_+^E$ satisfying $x(\delta(v))\leq b$ for each vertex $v\in V$. The maximum size of a $b$-matching in $G$ is defined as $\nu_b(G)$.
\end{definition}

\begin{definition} Define $\mathcal{N}_{r}$ as the set of servers that do not store any messages that are replicated fewer than $r$ times.
\begin{align}
\mathcal{N}_r&\triangleq \{n\in[N]\Big| m\in\mathcal{M}_n\implies \rho_m>r\}.
\end{align}
\end{definition}

\noindent It is  worthwhile to recall that from basic results in graph theory (see Chapter 30, Section 30.1 of Schrijver \cite{Schrijver}), it is  known that
\begin{align}
\nu_2(G)&=\min\{|V \backslash U|+|\mathcal{N}(U)|~\Big|~ U\subset V,\mbox{ and } U \mbox{ is a stable set}\}.\label{eq:nu2}
\end{align}

\noindent With this we are ready to state our final result.
\begin{theorem}\label{thm:GTPIR}
The asymptotic capacity of GTPIR with $\rho_m\leq T+2$ for all $m\in[M]$, i.e., when each message set is replicated no more than $(T+2)$ times, is
\begin{align}
C_\infty&=\left\{
\begin{array}{ll}
0,& \rho_{\min}\leq T\\
\frac{2}{\nu_{2}(G[\mathcal{N}_{T+2}])+2|\mathcal{N}_{T+1}|},& \rho_{\min}>T
\end{array}
\right..
\end{align}
\end{theorem}
The proof of Theorem \ref{thm:GTPIR} appears in Section \ref{proof:GTPIR}. While the converse bound for Theorem \ref{thm:GTPIR}  follows directly from the general converse bound in Theorem \ref{thm:converse}, the achievability goes beyond the scheme of Theorem \ref{thm:ach}, to involve a limited generalization to private computation that is presented in Section \ref{sec:GTPC}. As an interesting special case of Theorem \ref{thm:GTPIR}, note that if all messages are $T+2$ replicated, i.e., $\mathcal{N}_{T+1}$ is an empty set, then the asymptotic capacity is exactly $2/\nu_2(G)$.
\subsection{Examples}\label{sec:examples}
\begin{figure}[h]
\begin{center}
\begin{tikzpicture}
\node[circle, thick, draw=red,  inner sep = 2] (N1) at (0.75, 0.75) {\footnotesize $1$};
\node[circle, thick, draw=red,  inner sep = 2] (N2) at (0,0) {\footnotesize $2$};
\node[circle, thick, draw=red,  inner sep = 2] (N3) at (0.75,-0.75) {\footnotesize $3$};
\node[circle,  thick,  draw=red,  inner sep = 2] (N4) at (1.5,0) {\footnotesize $4$};
\node[circle, thick, fill=red!10, draw=red,  inner sep = 2] (N5) at (2.25,0.75) {\footnotesize $5$};
\node[circle, thick, fill=red!10, draw=red,  inner sep = 2] (N6) at (2.25,-0.75) {\footnotesize $6$};
\node[circle,   thick,   draw=black,  inner sep = 2] (N7) at (3,0) {\footnotesize $7$};
\node[circle,  thick, draw=black,  inner sep = 2] (N8) at (4,0) {\footnotesize $8$};
\draw [thick, red](N1)--(N2)--(N3)--(N1);
\draw [thick] (N1)--(N4)--(N3);
\draw[thick, red] (N4)--(N5);
\draw[thick, red] (N4)--(N6);
\draw [thick](N4)--(N7)--(N5);
\draw [thick] (N6)--(N7)--(N8);

\end{tikzpicture} 
\end{center}
\caption{The graph $G[V,E]$ for Example \ref{ex:Unotempty}.}\label{fig:gve}
\end{figure}
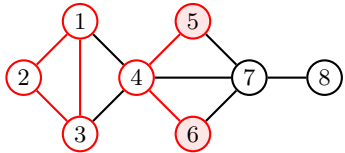

Let us consider a few  more examples to illustrate our results. For these examples we set $X=0,T=1$ for simplicity, but similar examples are easily constructed for  $X>0, T>1$ as well.
\begin{enumerate}
\item  \label{ex:symnocover1} Consider $M=4$ message sets, stored at $N=4$ servers according to the replication pattern $\mathcal{R}_1=(1,2,4)$, $\mathcal{R}_2=(1,2,3)$, $\mathcal{R}_3=(1,3,4)$. Since every message is $3$-replicated, according to Theorem \ref{thm:ach} we have $C_\infty\geq 2/4=1/2$. For the converse we note that $\mathcal{R}_1\implies D_1+D_2\geq 1$, $\mathcal{R}_2\implies D_2+D_3\geq 1$, $\mathcal{R}_3\implies D_3+D_4\geq 1, D_4+D_1\geq 1$, and adding these bounds gives us $D_1+D_2+D_3+D_4\geq 2$. Thus we have $C_\infty=1/2$ for this example. Note that this example does not contain an exact $b$-cover for any positive integer $b$, but the asymptotic capacity for this example is still  $C_\infty=(\rho_{\min}-X-T)/N$.
\item \label{ex:symnocover2} Consider $M=3$ message sets stored at $N=5$ servers according to the replication pattern $\mathcal{R}_1=(1,3,4), \mathcal{R}_2=(3,4,5), \mathcal{R}_3=(2,3,5)$, so that every message is $3$-replicated, but the storage is not  symmetric, nor does it contain an exact $b$-cover. For the converse we note that $\mathcal{R}_1\implies D_4+D_1\geq 1, D_1+D_3\geq 1$; $\mathcal{R}_3\implies D_3+D_2\geq 1, D_2+D_5\geq 1$; $\mathcal{R}_2\implies D_5+D_4\geq 1$; and combining these bounds gives us the converse bound as $C_\infty\leq \max_{\mathcal{D}} 1/(\sum_{n\in[5]}D_n)\geq 5/2$. Since $\rho_{\min}=3$, Theorem \ref{thm:ach} shows that the rate $(\rho_{\min}-X-T)/N=2/5$ is achievable, so that $C_\infty=2/5$ for this example.
\item Consider $M=3$ message sets stored at $N=5$ servers according to the replication pattern $\mathcal{R}_1=(1,3,4), \mathcal{R}_2=(1,3,4,5), \mathcal{R}_3=(2,3,5)$, so that messages in $\mathcal{W}_2$ are $4$-replicated while those in $\mathcal{W}_1, \mathcal{W}_3$ are only $3$-replicated. For the converse we note that $\mathcal{R}_1\implies D_1+D_3\geq 1, D_3+D_4\geq 1, D_4+D_1\geq 1$; while $\mathcal{R}_3\implies 2D_2+2D_5\geq 2$. Adding them up we have the bound $D_1+D_2+D_3+D_4+D_5\geq 5/2$, which gives us the converse bound $C_\infty\leq 2/5$. 
Since $\rho_{\min}=3$, the lower bound from Theorem \ref{thm:ach} is also $2/5$, so that $C_\infty=2/5$ for this example. Note that we could eliminate any one element from $\mathcal{R}_2$ so that messages in $\mathcal{W}_2$ are also only $3$-replicated, but that would not change the asymptotic capacity. Or we could add one more element to $\mathcal{R}_2$ so that messages in $\mathcal{W}_2$ are replicated at every server, and that would also not change the capacity. Thus, this example illustrates redundant storage. 

\item \label{ex:redserv} Consider $M=2$ message sets stored at $N=5$ servers according to the replication pattern $\mathcal{R}_1=(1,2,3,4)$, $\mathcal{R}_2=(2,3,4,5)$, so that each message is $4$-replicated. The converse from Theorem \ref{thm:converse} says $C_\infty\leq 2/3$, but since  $\rho_{\min}=4$,  Theorem \ref{thm:ach}  applied directly only proves the achievability of rate $(\rho_{\min}-X-T)/N=3/5$ which does not match the converse bound. However, note that if we eliminate Server $1$ and Server $5$, then we are left with the same\footnote{Note that while some servers may be eliminated (i.e., not used) by an achievable scheme, the message sets cannot be reduced because the achievable scheme must still work for all messages.} $M=2$ message sets stored at $N'=3$ servers according to the replication pattern $\mathcal{R}_1'=(2,3,4), \mathcal{R}_2'=(2,3,4)$, for which $\rho_{\min}'=3$, and Theorem \ref{thm:ach} shows that the rate $(\rho_{\min}'-X-T)/N'=2/3$ is achievable, which indeed matches the converse bound. Thus, the asymptotic capacity for this example is $C_\infty = 2/3$. The example shows that achievable rates may be improved by eliminating redundant servers.
\item Consider $M=4$ message sets stored at $N=5$ servers according to the storage pattern $\mathcal{R}_1=(1,2,3), \mathcal{R}_2=(2,3,4), \mathcal{R}_3=(1,3,5), \mathcal{R}_4=(2,4)$, so that messages in $\mathcal{W}_1, \mathcal{W}_2,\mathcal{W}_3$  are $3$-replicated, while messages in $\mathcal{R}_4$ are $2$-replicated, and $\rho_{\min}=2$. The achievable scheme from Theorem \ref{thm:ach} achieves a rate $1/5$, however Theorem \ref{thm:GTPIR} builds upon that scheme to achieve the rate $2/7$ which also matches the converse. Thus, for this setting, the capacity is settled by Theorem \ref{thm:GTPIR} as $C_\infty=2/7$.
\item \label{ex:Unotempty} Consider $M=5$ message sets stored at $N=8$ servers according to the storage pattern $\mathcal{R}_1=(1,2,3), \mathcal{R}_2=(1,3,4), \mathcal{R}_3=(4,5,7), \mathcal{R}_4=(4,6,7), \mathcal{R}_5=(7,8)$. The capacity for this case is  settled by Theorem \ref{thm:GTPIR} as $2/9$. To explicitly see the converse bound, note that in \eqref{eq:defD} $\mathcal{R}_1\implies D_1+D_2+D_3\geq 3/2$; $\mathcal{R}_5\implies D_7\geq 1, D_8\geq 1$; and $\mathcal{R}_3\implies D_4+D_5\geq 1$. Adding these bounds we have $D_1+D_2+D_3+D_4+D_5+D_7+D_8\geq 9/2$, which implies that asymptotically the total normalized download $D\geq 9/2$ and the converse bound follows. The graph representation for this setting, $G(V,E)$ is shown in Figure \ref{fig:gve}. Vertices in $\mathcal{N}_{3}=\{1,2,3,4,5,6\} $ are shown with a red border, while vertices in $\mathcal{N}_2=\{7,8\}$ are shown with a black border. The maximum size of a $2$-matching  on $G[\mathcal{N}_3]$ is $5$, corresponding to the $5$ edges shown in red. Alternatively, it corresponds to the choice of $U=\{5,6\}\subset\mathcal{N}_3$ in \eqref{eq:nu2}. Note that while $U$ has $2$ neighbors in $G$, i.e., $\mathcal{N}(U)=\{4,7\}$, it has only $1$ neighbor in $\mathcal{N}_3$,  i.e., $\mathcal{N}(U)\cap\mathcal{N}_3=\{4\}$. Therefore, $\nu_2(G[\mathcal{N}_3])+2|\mathcal{N}_2|=|\mathcal{N}_3\setminus U|+|\mathcal{N}(U)\cap\mathcal{N}_3|+2|\mathcal{N}_2|=4+1+2(2)=9$.  Achievability follows by the scheme  presented in the proof of Theorem \ref{thm:GTPIR},  downloading a symbol  from each of $[N]\setminus U=\{1,2,3,4,7,8\}$, and  downloading another symbol from each of $\mathcal{N}(U)\cup\mathcal{N}_2=\{4,7,8\}$ according to a private computation scheme described in Section \ref{sec:GTPC}, for a total download of $9$ symbols from which $2$ desired symbols are retrieved.
\end{enumerate}

\subsection{Solution Structure inspired by Dual GRS Codes}\label{sec:grsintuit}
The most interesting aspect of the achievable scheme in Theorem \ref{thm:ach} is  a generalized query and storage structure that is inspired by dual GRS codes. Since the storage and query structure for XSTPIR in \cite{Jia_Sun_Jafar_XSTPIR} was based on RS codes, the generalization to GRS code structure for GXSTPIR is somewhat serendipitous (note that the $G$ in GRS codes is not automatically associated with the $G$ in GXSTPIR which stands for Graph based replicated storage). It is also surprisingly effective, as explained intuitively in this section.

Before discussing how GRS codes are a part of the solution, let us illustrate the nature of the problem with a simple example. Let us consider a very basic setting, where we have $M=4$ subsets of messages,  $N=4$ servers, and $\forall m\in[M]$, we have $\mathcal{R}_m=[N]\setminus\{m\}$, i.e., messages in $\mathcal{W}_m$ are stored at all servers except Server $m$. Let $V_m, m\in[M]$ be four vectors in $\mathbb{F}$, each of size $N\times 1$, such that the vector $V_m$ has a zero in its $m^{th}$ coordinate (reflecting the fact that messages in $\mathcal{W}_m$ are not stored at Server $m$) and all other coordinates are non-zero. Then, as we will explain shortly, the rank of the matrix $[V_1, V_2, V_3, V_4]$ reflects the number of dimensions occupied by interference, i.e., downloaded symbols that are undesired. For example, suppose we are operating in $\mathbb{F}_5$ and we choose,
\begin{align}
V=[V_1, V_2, V_3, V_4]&=\left[
\begin{matrix}
0&1&1&1\\
1&0&3&2\\
1&2&0&4\\
1&3&1&0
\end{matrix}
\right]
\end{align}
which has rank $2$. Then this choice corresponds to a scheme where interference occupies $\mbox{rank}(V)=2$ out of the $N=4$ dimensions, leaving the remaining $2$ dimensions available for retrieving desired message symbols. To see this explicitly, suppose each message is comprised of $L=2$ symbols, $W_{m,k}=(W_{m,k}(1), W_{m,k}(2))$ in $\mathbb{F}_5$, and the user desires the message $W_{\mu,\kappa}\in\mathcal{W}_\mu$. The download from the $n^{th}$ server is the  $n^{th}$ row of the following $N\times 1$ vector.
\begin{align}
V&=\left(\sum_{k\in[K_1],\ell\in[L]}W_{1,k}(\ell)Z_{1,k,(\ell)}\right)V_1+\left(\sum_{k\in[K_2],\ell\in[L]}W_{2,k}(\ell)Z_{2,k,(\ell)}\right)V_2\notag\\
&\hspace{1cm}+\left(\sum_{k\in[K_3],\ell\in[L]}W_{3,k}(\ell)Z_{3,k,(\ell)}\right)V_3+
\left(\sum_{k\in[K_4],\ell\in[L]}W_{4,k}(\ell)Z_{4,k,(\ell)}\right)V_4\\
&\hspace{1cm}+{W}_{\mu,\kappa}(1)F^{[\mu,\kappa]}_{(1)}+{W}_{\mu,\kappa}(2)F^{[\mu,\kappa]}_{(2)}
\end{align}
The vectors $F^{[\mu,\kappa]}_{(1)}, F^{[\mu,\kappa]}_{(2)}$ are two $4\times 1$ vectors, called demand vectors that help retrieve the desired message symbols. To preserve privacy, the demand vectors $F^{[\mu,\kappa]}_{(1)}, F^{[\mu,\kappa]}_{(2)}$ must also have zeros in the coordinates where $V_\mu$ has zeros. 
The $Z_{k,m,(\ell)}$ random variables are i.i.d. uniform noise terms added to hide the demand vectors contained  in the query sent to each server, thus ensuring privacy of user's demand. The demand vectors, which carry the $2$ desired message symbols must be linearly independent of $V_1, V_2, V_3, V_4$ which carry only interference. To retrieve his desired message, the user projects $V$ into the $2$ dimensional null space of $V_1, V_2, V_3, V_4$, where all interference disappears and only the two desired signal dimensions remain, from which the $2$ desired symbols are retrieved. The rate achieved by this scheme is $2/4=1/2$ which is also the asymptotic capacity for this setting (converse follows from Theorem \ref{thm:converse}).

From this  example, it is clear that the problem is related to min-rank of the $V$ matrix subject to  constraints on which terms take zero or non-zero values. These constraints are affected not only by the given storage structure, but also from the possibility of redundant servers\footnote{As illustrated by examples in Section \ref{sec:examples} the solution may be further optimized on storage structure by ignoring  redundant storage.} as well as privacy and correctness constraints, e.g., because  demand vectors  must share the same structure to ensure privacy. Evidently, PIR with graph based storage is connected to other problems such as index coding, where also min-rank is important \cite{Birk_Kol}. For arbitrary storage patterns such min-rank problems can be difficult to solve in general. However, now let us consider what happens if every message is replicated the same number of times, $|\mathcal{R}_m|=\rho_m=\rho_{\min}$ for all $m\in[M]$. As will be shown in the proof of Theorem \ref{thm:GTPIR}, even if replication factors vary across messages, schemes for such settings may use the constant-replication-factor schemes as their essential building blocks. Thus, the constant-replication-factor setting is of fundamental significance. It is also the setting where we exploit the structure of dual GRS codes.

For simplicity we will only consider a setting with $X=0$ and $T=1$.
Consider such a setting with an arbitrary number of message sets $M$, with $N=5$ servers, constant-replication-factor $\rho_{min}=3$, and an arbitrary storage pattern reflected in the structure of the following $V$ matrix.
\begin{align}
V&=&\kbordermatrix{
&m=1&m=2&m=3&\cdots&m=M\\
\mbox{\tiny Server 1}&v_{1,1}&0&v_{3,1}&\cdots&v_{M,1}\\
\mbox{\tiny Server 2}&0&v_{2,2}&v_{3,2}&\cdots&0\\
\mbox{\tiny Server 3}&v_{1,3}&v_{2,3}&0&\cdots&v_{M,3}\\
\mbox{\tiny Server 4}&v_{1,4}&0&v_{3,4}&\cdots&0\\
\mbox{\tiny Server 5}&0&v_{2,5}&0&\cdots&v_{M,5}
}
\end{align}
Note that the $m^{th}$ column has exactly $\rho_m=3$ non-zero entries corresponding to the $3$ servers that store the messages in $\mathcal{W}_m$. The structure of each column is arbitrary, fixed by the given storage pattern, but each column must have exactly $3$ non-zero entries. For this setting, it turns out that regardless of the value of $M$, it is possible to choose non-zero values for $v_{m,n}$ such that the rank of this matrix is not more than $3$, i.e., all interference can be limited to $3$ dimensions. This is done as follows. Let $\beta_n$ be distinct non-zero constants for all $n\in[N]$. Furthermore, let us define,
\begin{align}
v_{m,n}&=\left(\prod_{n'\in\mathcal{R}_m\setminus\{n\}}(\beta_n-\beta_{n'})\right)^{-1}\label{eq:crucialex}
\end{align}
Based on dual GRS codes (see Lemma \ref{lemma:grs}), it turns out that this choice of $v_{m,n}$ ensures that
\begin{align}
\sum_{n\in\mathcal{R}_m}v_{m,n}\beta_n^{j}=0\label{eq:j}
\end{align}
for all $j\in\{0,1,\cdots,\rho_{\min}-2\}$. For this example, since $\rho_{\min}=3$, it means that $\sum_{n\in\mathcal{R}_m}v_{m,n}=0$, and $\sum_{n\in\mathcal{R}_m}v_{m,n}\beta_n=0$. Writing this out explicitly, we have
\begin{align}
&\left[\begin{matrix}
1&1&1&1&1\\
\beta_1&\beta_2&\beta_3&\beta_4&\beta_5
\end{matrix}
\right]
\left[\begin{matrix}
\frac{1}{(\beta_1-\beta_3)(\beta_1-\beta_4)}&0&\frac{1}{(\beta_1-\beta_2)(\beta_1-\beta_4)}&\cdots&\frac{1}{(\beta_1-\beta_3)(\beta_1-\beta_5)}\\
0&\frac{1}{(\beta_2-\beta_3)(\beta_2-\beta_5)}&\frac{1}{(\beta_2-\beta_1)(\beta_2-\beta_4)}&\cdots&0\\
\frac{1}{(\beta_3-\beta_1)(\beta_3-\beta_4)}&\frac{1}{(\beta_3-\beta_2)(\beta_3-\beta_5)}&0&\cdots&\frac{1}{(\beta_3-\beta_1)(\beta_3-\beta_5)}\\
\frac{1}{(\beta_4-\beta_1)(\beta_4-\beta_3)}&0&\frac{1}{(\beta_4-\beta_1)(\beta_4-\beta_2)}&\cdots&0\\
0&\frac{1}{(\beta_5-\beta_2)(\beta_5-\beta_3)}&0&\cdots&\frac{1}{(\beta_5-\beta_1)(\beta_5-\beta_3)}
\end{matrix}\right]
=\left[\begin{matrix}{\bf 0}\\{\bf 0}\end{matrix}\right]
\end{align}
which is easily verified because for any $n_1,n_2,n_3\in\mathcal[N]$,
\begin{align}
v_{m,n_1}+v_{m,n_2}+v_{m,n_3}&=\frac{(\beta_{n_2}-\beta_{n_3})+(\beta_{n_3}-\beta_{n_1})+(\beta_{n_1}-\beta_{n_2})}{(\beta_{n_1}-\beta_{n_2})(\beta_{n_1}-\beta_{n_3})(\beta_{n_2}-\beta_{n_3})}=0,\\
v_{m,n_1}\beta_{n_1}+v_{m,n_2}\beta_{n_2}+v_{m,n_3}\beta_{n_3}&=\frac{(\beta_{n_2}-\beta_{n_3})\beta_{n_1}+(\beta_{n_3}-\beta_{n_1})\beta_{n_2}+(\beta_{n_1}-\beta_{n_2})\beta_{n_3}}{(\beta_{n_1}-\beta_{n_2})(\beta_{n_1}-\beta_{n_3})(\beta_{n_2}-\beta_{n_3})}=0.
\end{align}

Thus, there are $\rho_{\min}-1=2$ vectors along which $V$ has null projection, corresponding to $j=0$ and $j=1$ in \eqref{eq:j}. These two interference free dimensions allow us to retrieve $2$ desired symbols, achieving a rate of $2/5$ for this example.  

As another example, consider a setting with an arbitrary number of messages $M$ and an arbitrary number of servers $N$, where each message is replicated $4$ times, i.e., $\rho_m=\rho_{\min}=4$ for all $m\in[M]$. Given an arbitrary $4$-replicated storage structure, choosing $v_{m,n}$ according to \eqref{eq:crucialex} allows us to find $\rho_{\min}-1=3$ dimensions along which interference is nulled, corresponding to $j=0, j=1$, and $j=2$ in \eqref{eq:j}. This is illustrated below.
\begin{align}
\left[\begin{matrix}
1&1&\cdots&1\\
\beta_1&\beta_2&\cdots&\beta_N\\
\beta_1^2&\beta_2^2&\cdots&\beta_N^2\\
\end{matrix}
\right]
\kbordermatrix{
&&\mbox{\tiny Column $m$}&\\
&\vdots&{\bf 0}&\vdots\\
\mbox{\tiny row $n_1$}&\cdots&v_{m,n_1}&\cdots\\
&\vdots&{\bf 0}&\vdots\\
\mbox{\tiny row $n_2$}&\cdots&v_{m,n_2}&\cdots\\
&\vdots&{\bf 0}&\vdots\\
\mbox{\tiny row $n_3$}&\cdots&v_{m,n_3}&\cdots\\
&\vdots&{\bf 0}&\vdots\\
\mbox{\tiny row $n_4$}&\cdots&v_{m,n_4}&\cdots\\
&\vdots&{\bf 0}&\vdots
}&=\left[\begin{matrix}{\bf 0}\\{\bf 0}\\{\bf 0}\end{matrix}\right].
\end{align}
Column $m$ corresponds to an arbitrary message set $\mathcal{W}_m$ that is replicated at the $4$ servers $n_1, n_2, n_3, n_4$, and it is easily verified that if $v_{m,n}$ are chosen according to \eqref{eq:crucialex} then
\begin{align}
v_{m,n_1}+v_{m,n_2}+v_{m,n_3}+v_{m,n_4}&=0,\\
\beta_{n_1}v_{m,n_1}+\beta_{n_2}v_{m,n_2}+\beta_{n_3}v_{m,n_3}+\beta_{n_4}v_{m,n_4}&=0,\\
\beta_{n_1}^2v_{m,n_1}+\beta_{n_2}^2v_{m,n_2}+\beta_{n_3}^2v_{m,n_3}+\beta_{n_4}^2v_{m,n_4}&=0.
\end{align}
Thus, there are $3$ interference-free dimensions which allow us to retrieve $3$ desired symbols for a rate of $3/N$.

In general, if the $V$ matrix has $\rho_{\min}$ non-zero entries in each column, then by assigning $v_{m,n}$ according to \eqref{eq:crucialex} there are $\rho_{\min}-1$ dimensions that are interference free, corresponding to $j\in\{0,1,\cdots,\rho_{\min}-2\}$ in \eqref{eq:j}, along which $\rho_{\min}-1$ desired symbols  can be retrieved to achieve the rate $(\rho_{\min}-1)/N$, which matches $(\rho_{\min}-X-T)/N$ for $X=0, T=1$. When $T>1$ and/or $X>0$, then additional interference terms enter into the picture due to the additional noise terms needed to protect the messages ($X$-security) and the queries ($T$-privacy). Following the construction previously introduced for XSTPIR, these additional interference dimensions are restricted by using cross-subspace alignment \cite{Jia_Sun_Jafar_XSTPIR}. Fortunately, since the storage and query structure used for XSTPIR in \cite{Jia_Sun_Jafar_XSTPIR} is also based on Reed Solomon Codes, it turns out to be compatible with the additional structure imposed by the choice of $v_{m,n}$ in \eqref{eq:crucialex} according to dual Generalized Reed Solomon Codes. Combining both ideas, it turns out that the number of interference free dimensions that remain available for desired message symbols is equal to $\rho_{\min}-X-T$, which allows us to achieve a rate of $(\rho_{\min}-X-T)/N$. The details are left to the proof of Theorem \ref{thm:ach}.

\section{Proof of Theorem \ref{thm:ach}}\label{proof:ach}
In this section we present the achievable scheme for GXSTPIR for arbitrary $N,T, X, M, K_m,\rho_m$ values that allows private retrieval of any desired message at a rate $R=\frac{\rho_{\min}-X-T}{N}$. Without loss of generality we will assume that $\rho_m=\rho_{\min}$ for all $m\in[M]$. For any message that is replicated more than $\rho_{\min}$ times, the scheme can be applied by arbitrarily choosing any $\rho_{\min}$ replications of that message and ignoring the rest. In order to achieve the rate $R=\frac{\rho_{\min}-X-T}{N}$, the scheme will retrieve $\rho_{\min}-X-T$ desired symbols by downloading one symbol from each server. 

The scheme operates over a block where each message is comprised of 
$L$ symbols and we have 
\begin{align}
L=\rho_{\min}-X-T.
\end{align}
All symbols are in $\mathbb{F}_q$ and without loss of generality we will assume that $q> N+L$.
Let $\beta_{[N]}$ be distinct non-zero values in $\mathbb{F}_q$ such that
\begin{align}
\beta_n+\ell&\neq 0, &&\forall n\in[N], \ell\in[L].
\end{align}
Such $\beta_n$ must exist because $q> L+N$.
Server $n$ stores,
\begin{align}
S_n&=\{{\bf W}_{m,(1)}^{(n)}, {\bf W}_{m,(2)}^{(n)}, \cdots, {\bf W}_{m,(L)}^{(n)}, \forall m\in\mathcal{M}_n\}\\
{\bf W}_{m,(\ell)}^{(n)}&={\bf W}_{m,(\ell)}+\sum_{x\in[X]}(\ell+\beta_n)^x{\bf Z}_{m,x,(\ell)}\\
{\bf W}_{m,(\ell)}&=[W_{m,1}(\ell),W_{m,2}(\ell),\cdots,W_{m,K_m}(\ell) ], &&\forall \ell\in[L].
\end{align}
Thus, for all $m\in[M]$, the $1\times K_m$ row vector ${\bf W}_{m,(\ell)}$ contains the $\ell^{th}$ symbol from every message in $\mathcal{W}_m$.
For all $m\in[M], x\in[X], \ell\in[L]$, the $1\times K_m$ row vectors  ${\bf Z}_{m,x,(\ell)}$ are comprised of i.i.d. uniform noise symbols. Any  message symbol $W_{m,k}(\ell)$ that is secret-shared among servers $\mathcal{R}_m$,  is protected by the $X$ noise symbols ${\bf Z}_{m,1,(\ell)}(k), {\bf Z}_{m,2,(\ell)}(k), \cdots, {\bf Z}_{m,X,(\ell)}(k)$ that are i.i.d. uniform and coded according to an MDS($X,\rho_{\min}$)  code, so that the shares accessible to any set of up to $X$ colluding servers are independent of $W_{m,k}(\ell)$. Thus the scheme is $X$-secure. 

The query sent to Server $n$ is
\begin{align}
Q_n^{[\mu,\kappa]}&=\{{\bf Q}_{m,n,(\ell)}^{[\mu,\kappa]},\forall m\in\mathcal{M}_n, \ell\in[L]\}\\
\intertext{where,}
{\bf Q}_{m,n,(\ell)}^{[\mu,\kappa]}&=\frac{v_{m,n}}{\ell+\beta_n}\left({\bf F}_{m}^{[\mu,\kappa]}+\sum_{t\in[T]}(\ell+\beta_n)^t{\bf Z}'_{m,t,(\ell)}\right)\label{eq:vmn}
\end{align}
${\bf F}_{m}^{[\mu,\kappa]}$ are demand vectors defined as
\begin{align}
{\bf F}_{m}^{[\mu,\kappa]}&=\left\{
\begin{array}{ll}
{\bf e}_\kappa, & \mbox{if } m=\mu,\\
{\bf 0}, &\mbox{otherwise.}\label{eq:chooseFm}
\end{array}
\right.
\end{align}
where ${\bf e}_\kappa$ is the $\kappa^{th}$ column of the $K_m\times K_m$ identity matrix. 
The values of ${\bf F}_m^{[\mu,\kappa]}$ are kept private from any set of up to $T$ colluding servers, by the $K_m\times 1$ column vectors ${\bf Z}'_{m,t,(\ell)}$   comprised of i.i.d uniform noise symbols, for all $m\in[M], t\in[T], \ell\in[L]$. Note that the noise vectors that protect ${\bf F}_{m}^{[\mu,\kappa]}$ are coded according to an MDS($T,\rho_{\min}$) code spread across the queries sent to servers in $\mathcal{R}_m$, i.e., all queries that contain ${\bf F}_{m}^{[\mu,\kappa]}$, so that the queries accessible to any set of up to $T$ servers reveal no information about the demand vectors. Thus, the scheme is $T$-private.

The constant values $v_{m,n}$ in \eqref{eq:vmn} are defined as
\begin{align}
v_{m,n}&\triangleq \left(\prod_{n'\in\mathcal{R}_m\setminus\{n\}}(\beta_n-\beta_{n'})\right)^{-1}
\end{align}
As shown in Lemma \ref{lemma:grs} in Appendix \ref{app:lemmas} using the properties of dual GRS codes, this choice of $v_{m,n}$ satisfies the crucial property that
\begin{align}
\sum_{n\in\mathcal{R}_m}v_{m,n}\beta_n^j&=0\label{eq:crucialx}
\end{align}
for all $m\in[M]$ and for all $j\in\{0,1,\cdots,\rho_{\min}-2\}$. 

The answer returned by Server $n$ is
\begin{align}
A_n^{[\mu,\kappa]}&=\sum_{\ell\in[L]}\sum_{m\in\mathcal{M}_n}{\bf W}_{m,(\ell)}^{(n)}{\bf Q}_{m,n,(\ell)}^{[\mu,\kappa]}
\end{align}
Upon receiving all $N$ answers, the user evaluates the $L$ values $Y_1, Y_2, \cdots, Y_L$, as follows.
\begin{align}
\left[
\begin{matrix}
Y_1\\
Y_2\\
\vdots\\
Y_L
\end{matrix}
\right]&=\left[
\begin{matrix}
1& 1& \cdots& 1\\
\beta_1&\beta_2&\cdots&\beta_N\\
\vdots&\vdots&\cdots&\vdots\\
\beta_1^{L-1}&\beta_2^{L-1}&\cdots&\beta_N^{L-1}
\end{matrix}
\right]\left[
\begin{matrix}
A_1^{[\mu,\kappa]}\\
A_2^{[\mu,\kappa]}\\
\vdots\\
A_N^{[\mu,\kappa]}
\end{matrix}
\right]
\end{align}
so that for all $i\in[L]$,
\begin{align}
Y_i&=\sum_{n\in[N]}\beta_n^{i-1}A_n^{[\mu,\kappa]}\\
&=\sum_{n\in[N]}\beta_n^{i-1}\sum_{l\in[L]}\sum_{m\in\mathcal{M}_n}{\bf W}_{m,(\ell)}^{(n)}{\bf Q}_{m,n,(\ell)}^{[\mu,\kappa]}\\
&=\sum_{\ell\in[L]}\sum_{m\in[M]}\sum_{n\in\mathcal{R}_m}\beta_n^{i-1}{\bf W}_{m,(\ell)}^{(n)}{\bf Q}_{m,n,(\ell)}^{[\mu,\kappa]}\\
&=\sum_{\ell\in[L]}\sum_{m\in[M]}\sum_{n\in\mathcal{R}_m}\frac{v_{m,n}\beta_n^{i-1}}{\ell+\beta_n}\left({\bf W}_{m,(\ell)}+\sum_{x\in[X]}(\ell+\beta_n)^x{\bf Z}_{m,x,(\ell)}\right)\left({\bf F}_{m}^{[\mu,\kappa]}+\sum_{t\in[T]}(\ell+\beta_n)^t{\bf Z}'_{m,t,(\ell)}\right)\\
&=\sum_{\ell\in[L]}\sum_{m\in[M]}\sum_{n\in\mathcal{R}_m}\left(\frac{v_{m,n}\beta_n^{i-1}}{\ell+\beta_n}{\bf W}_{m,(\ell)}{\bf F}_{m}^{[\mu,\kappa]}+\sum_{t\in[T]}v_{m,n}\beta_n^{i-1}(\ell+\beta_n)^{t-1}{\bf W}_{m,(\ell)}{\bf Z}'_{m,t,(\ell)}\right.\notag\\
&\hspace{5cm}\left.+\sum_{x\in[X]}v_{m,n}\beta_n^{i-1}(\ell+\beta_n)^{x-1}{\bf Z}_{m,x,(\ell)}{\bf F}_{m}^{[\mu,\kappa]}\right.\notag\\
&\hspace{5cm}\left.+\sum_{x\in[X]}\sum_{t\in[T]}v_{m,n}\beta_n^{i-1}(\ell+\beta_n)^{x+t-1}{\bf Z}_{m,x,(\ell)}{\bf Z}'_{m,t,(\ell)}\right)\\
&=\sum_{\ell\in[L]}\sum_{m\in[M]}\sum_{n\in\mathcal{R}_m}\left(\frac{v_{m,n}\beta_n^{i-1}}{\ell+\beta_n}{\bf W}_{m,(\ell)}{\bf F}_{m}^{[\mu,\kappa]}\right)\notag\\
&\hspace{3cm}+\sum_{\ell\in[L]}\sum_{m\in[M]}\left(\sum_{t\in[T]}{\bf W}_{m,(\ell)}{\bf Z}'_{m,t,(\ell)}\left(\sum_{n\in\mathcal{R}_m}v_{m,n}\beta_n^{i-1}(\ell+\beta_n)^{t-1}\right)\right)\notag\\
&\hspace{3cm}+\sum_{\ell\in[L]}\sum_{m\in[M]}\left(\sum_{x\in[X]}{\bf Z}_{m,x,(\ell)}{\bf F}_{m}^{[\mu,\kappa]}\left(\sum_{n\in\mathcal{R}_m}v_{m,n}\beta_n^{i-1}(\ell+\beta_n)^{x-1}\right)\right)\notag\\
&\hspace{3cm}+\sum_{\ell\in[L]}\sum_{m\in[M]}\left(\sum_{x\in[X]}\sum_{t\in[T]}{\bf Z}_{m,x,(\ell)}{\bf Z}'_{m,t,(\ell)}\left(\sum_{n\in\mathcal{R}_m}v_{m,n}\beta_n^{i-1}(\ell+\beta_n)^{x+t-1}\right)\right)
\end{align}
The terms $\left(\sum_{n\in\mathcal{R}_m}v_{m,n}\beta_n^{i-1}(\ell+\beta_n)^{t-1}\right)$, $\left(\sum_{n\in\mathcal{R}_m}v_{m,n}\beta_n^{i-1}(\ell+\beta_n)^{x-1}\right)$ and \linebreak
$\left(\sum_{n\in\mathcal{R}_m}v_{m,n}\beta_n^{i-1}(\ell+\beta_n)^{x+t-1}\right)$ are  equal to zero because of \eqref{eq:crucialx}. This is because all of these can be expanded into  weighted sums of terms of the form $\sum_{n\in\mathcal{R}_m}v_{m,n}\beta_n^{j}$ for $j$ taking values in $\{0,1,\cdots, \rho_{\min}-2\}$. Let us show this explicitly for $\sum_{n\in\mathcal{R}_m}v_{m,n}\beta_n^{i-1}(\ell+\beta_n)^{t-1}$ as follows,
\begin{align}
\sum_{n\in\mathcal{R}_m}v_{m,n}\beta_n^{i-1}(\ell+\beta_n)^{t-1}&=\sum_{n\in\mathcal{R}_m}v_{m,n}\beta_n^{i-1}\left(\sum_{\tau\in\{0,1,\cdots,t-1\}}\binom{t-1}{\tau}\beta_n^\tau\ell^{t-1-\tau}\right)\\
&=\sum_{\tau\in\{0,1,\cdots,t-1\}}\binom{t-1}{\tau}\ell^{t-1-\tau}\left(\sum_{n\in\mathcal{R}_m}v_{m,n}\beta_n^{i+\tau-1}\right)\\
&=0
\end{align}
because $0\leq i+\tau-1\leq L+(T-1)-1=\rho_{\min}-X-2\leq \rho_{\min}-2$. 
It can be similarly shown that $\left(\sum_{n\in\mathcal{R}_m}v_{m,n}\beta_n^{i-1}(\ell+\beta_n)^{x-1}\right)=0$ and $\left(\sum_{n\in\mathcal{R}_m}v_{m,n}\beta_n^{i-1}(\ell+\beta_n)^{x+t-1}\right)=0$. Thus, we have,
\begin{align}
Y_i&=\sum_{\ell\in[L]}\sum_{m\in[M]}\sum_{n\in\mathcal{R}_m}\left(\frac{v_{m,n}\beta_n^{i-1}}{\ell+\beta_n}{\bf W}_{m,(\ell)}{\bf F}_{m}^{[\mu,\kappa]}\right)\\
&=\sum_{\ell\in[L]}\sum_{m\in[M]}{\bf W}_{m,(\ell)}{\bf F}_{m}^{[\mu,\kappa]}\left(\sum_{n\in\mathcal{R}_m}\frac{v_{m,n}\beta_n^{i-1}}{\ell+\beta_n}\right)\label{eq:switchlambda}\\
&=\sum_{\ell\in[L]}{\bf W}_{\mu,(\ell)}{\bf e}_\kappa\left(\sum_{n\in\mathcal{R}_\mu}\frac{v_{\mu,n}\beta_n^{i-1}}{\ell+\beta_n}\right)\label{eq:pluginF}\\
&=\sum_{\ell\in[L]}\sum_{n\in\mathcal{R}_\mu}W_{\mu,\kappa}(\ell)\frac{v_{\mu,n}\beta_n^{i-1}}{\ell+\beta_n}
\end{align}
Note that we used \eqref{eq:chooseFm} to obtain \eqref{eq:pluginF}. In matrix notation, we have,
\begin{align}
\left[
\begin{matrix}
Y_1\\
Y_2\\
\vdots\\
Y_L
\end{matrix}
\right]&=
\underbrace{\left[
\begin{matrix}
1& \cdots& 1\\
\beta_{\mathcal{R}_{\mu}(1)}&\cdots&\beta_{\mathcal{R}_{\mu}(\rho_m)}\\
\vdots&\vdots&\vdots\\
\beta_{\mathcal{R}_{\mu}(1)}^{L-1}&\cdots&\beta_{\mathcal{R}_{\mu}(\rho_m)}^{L-1}
\end{matrix}
\right]}_{A}\underbrace{\left[
\begin{matrix}
\frac{v_{\mu,\mathcal{R}_{\mu}(1)}}{1+\beta_{\mathcal{R}_{\mu}(1)}}&\dots&\frac{v_{\mu,\mathcal{R}_{\mu}(1)}}{L+\beta_{\mathcal{R}_{\mu}(1)}}\\
\vdots&\vdots&\vdots\\
\frac{v_{\mu,\mathcal{R}_{\mu}(\rho_m)}}{1+\beta_{\mathcal{R}_{\mu}(\rho_m)}}&\dots&\frac{v_{\mu,\mathcal{R}_{\mu}(\rho_m)}}{L+\beta_{\mathcal{R}_{\mu}(\rho_m)}}
\end{matrix}
\right]}_{B}
\left[
\begin{matrix}
W_{\mu,\kappa}(1)\\
W_{\mu,\kappa}(2)\\
\vdots\\
W_{\mu,\kappa}(L)
\end{matrix}
\right].
\end{align}
If the $L\times L$ matrix $AB$ is invertible, the desired message is retrievable. This can be proved as follows. Guaranteed by Lemma \ref{lemma:grs} and the definitions of $v_{m,n}$ and $\beta_{n}$, $\forall m\in[M], n\in[N]$, the rows of the $L\times\rho_m$ matrix $A$ generate the null space of the following $\rho_m\times (\rho_m-L)$ matrix. \begin{align}
C=\left[
\begin{matrix}
v_{\mu,\mathcal{R}_{\mu}(1)}&v_{\mu,\mathcal{R}_{\mu}(1)}\beta_{\mathcal{R}_{\mu}(1)}&\dots&v_{\mu,\mathcal{R}_{\mu}(1)}\beta_{\mathcal{R}_{\mu}(1)}^{\rho_m-L-1}\\
\vdots&\vdots&\vdots&\vdots\\
v_{\mu,\mathcal{R}_{\mu}(\rho_m)}&v_{\mu,\mathcal{R}_{\mu}(\rho_m)}\beta_{\mathcal{R}_{\mu}(\rho_m)}&\dots&v_{\mu,\mathcal{R}_{\mu}(\rho_m)\beta_{\mathcal{R}_{\mu}(\rho_m)}^{\rho_m-L-1}}
\end{matrix}
\right]
\end{align}
Next we note that by Lemma 5 in \cite{Jia_Sun_Jafar_XSTPIR}, the $\rho_m\times \rho_m$ matrix $[B|C]$ is invertible. 
Therefore the matrix $AB$ must be invertible, and the desired message is retrievable. Thus the scheme is correct. This completes the proof of Theorem \ref{thm:ach}. $\hfill\square$

\subsection{A Private Computation Scheme for $X=0$, $\rho_{\min}=T+1$.}\label{sec:GTPC}
From the description of the scheme, it is evident that the demand vectors are protected by the uniform noise, regardless of how they are chosen. Modifying the choice of demand vectors would allow the user to privately retrieve various forms of desired information, generalizing the scheme to broader applications. Here we present a simple example that will also be useful for the proof of Theorem \ref{thm:GTPIR}. 

Suppose there are no security constraints $(X=0)$ and every message is replicated $T+1$ times ($\rho_{\min}=T+1$), so that that our scheme operates over blocks comprised of $L=\rho_{\min}-X-T=1$ symbol per message. Recall that our scheme allows the user to retrieve an arbitrary message $W_{\mu,\kappa}$ at the rate $R=({\rho_{\min}-X-T})/{N}=1/N$ in this setting. Now, suppose instead of an arbitrary message, the user wants to retrieve  an arbitrary linear combination of all messages,
\begin{align}
\lambda(\mathcal{W})&\triangleq \sum_{m\in[M]}\sum_{k\in K_m}\lambda_{m,k}W_{m,k}(1)=\sum_{m\in[M]}{\bf W}_{m,(1)}{\bm\lambda}_m,&&\forall\ell\in[L]\\
\intertext{where}
{\bm \lambda}_m&=[\lambda_{m,1}, \lambda_{m,2},\cdots,\lambda_{m,K_m}]^T\in\mathbb{F}_q^{K_m\times 1}, &&\forall m\in[M],
\end{align}
are the combining coefficients  to be kept private from any set of up to $T$ colluding servers. This is a form of the private linear computation problem studied in \cite{Sun_Jafar_PC} applied here to  graph based replicated storage. To apply our scheme to this setting, replace the demand vectors ${\bf F}_{m}^{[\mu,\kappa]}$ with ${\bf F}_{m}^{[\lambda]}$ defined as follows.
\begin{align}
{\bf F}_{m}^{[\lambda]}&=\left(\sum_{n\in\mathcal{R}_m}\frac{v_{m,n}}{1+\beta_n}\right)^{-1}{\bm\lambda}_m
\end{align}
so that continuing from \eqref{eq:switchlambda} we have
\begin{align}
Y_i&=\sum_{\ell\in[L]}\sum_{m\in[M]}{\bf W}_{m,(\ell)}{\bf F}_{m}^{[\lambda]}\left(\sum_{n\in\mathcal{R}_m}\frac{v_{m,n}\beta_n^{i-1}}{\ell+\beta_n}\right),~~i\in[L]=\{1\}\\\\
\Rightarrow Y_1&=\sum_{m\in[M]}{\bf W}_{m,(1)}{\bm\lambda}_m\left(\sum_{n\in\mathcal{R}_m}\frac{v_{m,n}}{1+\beta_n}\right)^{-1}\left(\sum_{n\in\mathcal{R}_m}\frac{v_{m,n}}{1+\beta_n}\right)\\
&=\sum_{m\in[M]}{\bf W}_{m,(1)}{\bm\lambda}_m=\lambda(\mathcal{W})
\end{align}
Thus, a private computation scheme is readily obtained for the case where all messages are replicated at least $T+1$ times. The rate of this scheme is $(\rho_{\min}-T)/N=1/N$. Just as in \cite{Sun_Jafar_PC}, there is no rate loss relative to the case where the user wants to retrieve only one message $W_{\mu,\kappa}$.

\section{Proof of Theorem \ref{thm:converse}}\label{proof:converse}
Let $\mathcal{T}$ be a subset of $\mathcal{R}_m$, such that $|\mathcal{T}|=\max(|\mathcal{R}_m|, T)$. Let $\mathcal{X}$ be a subset of $\mathcal{R}_m\setminus\mathcal{T}$, such that $|\mathcal{X}|=\max(|\mathcal{R}_m|-|\mathcal{T}|, X)$. Note that it follows from the definition that $\mathcal{T}\cap\mathcal{X}=\emptyset$. From the decodability of message $W_{m,k}$ we have,
\begin{align}
L&= I\left(W_{m,k}~;~A_{[N]}^{[m,k]}\mid Q_{[N]}^{[m,k]}\right)\\
&\leq  I\left(W_{m,k}~;~A_{\mathcal{R}_m\setminus\mathcal{X}}^{[m,k]}, S_{[N]\backslash\mathcal{R}_m}, S_{\mathcal{X}}\mid Q^{[m,k]}_{[N]}\right)\label{eq:con1eq1}\\
&=  I\left(W_{m,k}~;~S_{[N]\backslash\mathcal{R}_m},S_{\mathcal{X}}\mid Q^{[m,k]}_{[N]}\right)+I\left(W_{m,k}~;~A_{\mathcal{R}_m\setminus\mathcal{X}}^{[m,k]}\mid S_{[N]\backslash\mathcal{R}_m}, S_{\mathcal{X}}, Q^{[m,k]}_{[N]}\right)\label{eq:con1eq2}\\
&=I\left(W_{m,k}~;~A_{\mathcal{R}_m\setminus\mathcal{X}}^{[m,k]}\mid S_{\mathcal{X}}, S_{[N]\backslash\mathcal{R}_m}, Q^{[m,k]}_{[N]}\right)\label{eq:con1eq3}\\
&=I\left(W_{m,k}~;~ A_{\mathcal{T}}^{[m,k]}, A_{(\mathcal{R}_m\setminus\mathcal{X})\backslash\mathcal{T}}^{[m,k]}\mid S_{\mathcal{X}}, S_{[N]\backslash\mathcal{R}_m}, Q^{[m,k]}_{[N]}\right)\label{eq:con1eq4}\\
&=I\left(W_{m,k}; A_{\mathcal{T}}^{[m,k]}\mid S_{\mathcal{X}}, S_{[N]\backslash\mathcal{R}_m}, Q^{[m,k]}_{[N]}\right)+I\left(W_{m,k}; A_{(\mathcal{R}_m\setminus\mathcal{X})\backslash\mathcal{T}}^{[m,k]}\mid A_{\mathcal{T}}^{[m,k]}, S_{\mathcal{X}}, S_{[N]\backslash\mathcal{R}_m}, Q^{[m,k]}_{[N]}\right)\label{eq:con1eq5}\\
&\leq I(W_{m,k}; A_{\mathcal{T}}^{[m,k]}\mid S_{\mathcal{X}}, S_{[N]\backslash\mathcal{R}_m}, Q^{[m,k]}_{[N]})+\sum_{n\in(\mathcal{R}_m\setminus\mathcal{X})\setminus\mathcal{T}}H(A_n^{[m,k]})\label{eq:con1eq6}\\
&\leq I(W_{m,k}; A_{\mathcal{T}}^{[m,k]}\mid S_{\mathcal{X}}, S_{[N]\backslash\mathcal{R}_m}, Q^{[m,k]}_{\mathcal{T}})+\sum_{n\in(\mathcal{R}_m\setminus\mathcal{X})\backslash\mathcal{T}}H(A_n^{[m,k]})\label{eq:con1eq7}\\
&\leq I(W_{m,k}; A_{\mathcal{T}}^{[m,k']}\mid S_{\mathcal{X}}, S_{[N]\backslash\mathcal{R}_m}, Q^{[m,k']}_{\mathcal{T}})+\sum_{n\in(\mathcal{R}_m\setminus\mathcal{X})\backslash\mathcal{T}}H(A_n^{[m,k']})\label{eq:con1eq8}
\end{align}
In \eqref{eq:con1eq1} we used the fact that $A_{[N]\backslash(\mathcal{R}_m\setminus\mathcal{X})}^{[m,k]}$ is a function of $\left(S_{[N]\backslash\mathcal{R}_m},S_{\mathcal{X}}, Q_{[N]}^{[m,k]}\right)$, and $I(A;f(B,C)\mid C)\leq I(A;f(B,C),B\mid C)=I(A;B\mid C)+I(A;f(B,C)\mid B,C)=I(A;B\mid C)$ where $f(B,C)$ is some function of $B,C$. The chain rule of mutual information is used for \eqref{eq:con1eq2}. For \eqref{eq:con1eq3} we used the fact that $(S_{[N]\backslash\mathcal{R}_m},S_{\mathcal{X}})$ is independent of $\left(W_{m,k}, Q_{[N]}^{[m,k]}\right)$ according to Lemma \ref{lemma:indSWq}. The next step, \eqref{eq:con1eq4} simply re-writes the same expression in different notation, while \eqref{eq:con1eq5} follows from chain rule of mutual information. For \eqref{eq:con1eq6} we used the fact that $I(A;B\mid C)=H(B\mid C)-H(B\mid A,C)\leq H(B)$ because entropy is non-negative and conditioning reduces entropy. 
 \eqref{eq:con1eq7} follows from Lemma \ref{lemma:NtoT}.   \eqref{eq:con1eq8} follows because $I(Q_{\mathcal{T}}^{[m,\kappa]}, A_{\mathcal{T}}^{[m,\kappa]}, S_{[N]}; \kappa)=0$ according to Lemma \ref{lemma:condpriv}. Equivalently, 
\begin{align}
\left(Q_{\mathcal{T}}^{[m,k]}, A_{\mathcal{T}}^{[m,k]}, S_{[N]}\right)&\sim
\left(Q_{\mathcal{T}}^{[m,k']}, A_{\mathcal{T}}^{[m,k']}, S_{[N]}\right)
\end{align}
for all $m\in[M]$ and $k,k'\in[K_m]$, which in turn implies \eqref{eq:con1eq8}. 

\noindent Summing \eqref{eq:con1eq8} over all $k\in[K_m]$ we have
\begin{align}
K_mL&\leq \left(\sum_{k\in[K_m]} I(W_{m,k}; A_{\mathcal{T}}^{[m,k']}\mid S_{\mathcal{X}}, S_{[N]\backslash\mathcal{R}_m}, Q^{[m,k']}_{\mathcal{T}})\right)+K_m\sum_{n\in(\mathcal{R}_m\setminus\mathcal{X})\backslash\mathcal{T}}H(A_n^{[m,k']})\label{eq:con1ceq1}\\
&\leq I(W_{m,1}, \cdots, W_{m,K_m}; A_{\mathcal{T}}^{[m,k']}\mid S_{\mathcal{X}}, S_{[N]\backslash\mathcal{R}_m}, Q^{[m,k']}_{\mathcal{T}})+K_m\sum_{n\in(\mathcal{R}_m\setminus\mathcal{X})\backslash\mathcal{T}}H(A_n^{[m,k']})\label{eq:con1ceq2}\\
&\leq H(A_{\mathcal{T}}^{[m,k']})+K_m\sum_{n\in(\mathcal{R}_m\setminus\mathcal{X})\backslash\mathcal{T}}H(A_n^{[m,k']})\label{eq:con1ceq3}
\end{align}
\eqref{eq:con1ceq2} follows from the chain rule of mutual information and repeated use of the property that $I(A;C\mid D)+I(B;C\mid D)\leq I(A;C\mid D)+I(B;C\mid A,D)=I(A,B;C\mid D)$ when $A,B$ are independent conditioned on $D$, i.e., $I(A;B\mid D)=0$. This conditional independence property for \eqref{eq:con1ceq2} is proved in Lemma \ref{lemma:condind}.  \eqref{eq:con1ceq3} follows from the facts that entropy is non-negative and conditioning reduces entropy, i.e., $I(A;B\mid C)=H(A\mid C)-H(A\mid B,C)\leq H(A\mid C)\leq H(A)$.

From \eqref{eq:con1ceq3}  we note that if $|\mathcal{R}_m|\leq X+T$ then $\mathcal{R}_m\setminus\mathcal{X}\backslash\mathcal{T}=\emptyset$, which means that as $K_m\rightarrow\infty$, we must have $H(A_\mathcal{T}^{[m,k']})\rightarrow\infty$, and since the download approaches infinity, the asymptotic capacity is zero.  This is the degenerate case in Theorem \ref{thm:converse}.

Having dealt with the degenerate setting, henceforth, let us assume that $|\mathcal{R}_m|>X+T$ for all $m\in[M]$. Since the capacity for this case is not zero (follows from achievability), there is no loss of generality in assuming that the asymptotic value of download cost is bounded, i.e., $H(A_{n}^{[m,k']})/K_m=o(1)$ as a function of $K_m$ for all $n\in[N]$. Recall that $f(x)=o(1)$ is equivalent to the condition that $\lim_{x\rightarrow\infty} f(x)=0$. In this case we have
\begin{align}
\sum_{n\in(\mathcal{R}_m\setminus\mathcal{X})\backslash\mathcal{T}}\frac{H(A_n^{[m,k']})}{L}+o(1)&\geq 1\\
\Rightarrow \sum_{n\in(\mathcal{R}_m\setminus\mathcal{X})\backslash\mathcal{T}}D_n+o(1)&\geq 1.\label{eq:Dbound}
\end{align}
where $D_n=\frac{H(A_n^{[m,k']})}{L}$ is defined as the value of download from server $n$, normalized by $L$. As $K\rightarrow\infty$ all $o(1)$ terms approach $0$ and we obtain the set of conditions that define $\mathcal{D}$ in \eqref{eq:defD}. The capacity bound in Theorem \ref{thm:converse}  for the non-degenerate setting follows from the definition of capacity as the supremum of $L/D=(D_1+\cdots+D_N)^{-1}$. $\hfill\square$

\section{Proof of Theorem \ref{thm:GTPIR}}\label{proof:GTPIR}
\subsection{Proof of Converse for Theorem \ref{thm:GTPIR}}
It already follows from Theorem \ref{thm:converse} that if $\rho_{\min}\leq T$ then the capacity is zero. So let us assume that $\rho_{\min}>T$. Theorem \ref{thm:GTPIR} also limits $\rho_m\leq T+2$ for all $m\in[M]$, therefore we must have $\rho_m\in\{T+1, T+2\}$ for all $m\in[M]$, i.e., every message is either $(T+1)$-replicated or $(T+2)$-replicated. Recall that $\mathcal{N}_{T+2}$ is the set of servers that do not store any messages that are $(T+1)$-replicated. The remaining servers are in $\mathcal{N}_{T+1}$.

According to the general converse bound in Theorem \ref{thm:converse}, the asymptotic capacity $C_\infty$ is bounded above by the maximum value of $(D_1+\cdots+D_N)^{-1}$ subject to the constraints,
\begin{align}
 D_u+D_v\geq 1, &&\forall uv\in E[\mathcal{N}_{T+2}]
\label{eq:duv}\\
 D_t\geq 1,&&\forall t\in[\mathcal{N}_{T+1}]\label{eq:dt}
\end{align}
We use the notation $G[\mathcal{N}_{T+2}]$ to represent the induced subgraph of $G[V,E]$ whose vertex set is $\mathcal{N}_{T+2}$ and whose edge set, denoted $E[\mathcal{N}_{T+2}]$ consists of all edges $uv\in E$ such that $u,v\in\mathcal{N}_{T+2}$.  Recall that a $2$-matching in $G[\mathcal{N}_{T+2}]$ is  a vector $x$ that assigns to each edge $uv\in E[\mathcal{N}_{T+2}]$, a value from $\{0,1,2\}$ such that the sum of values assigned to all edges in $E[\mathcal{N}_{T+2}]$ that are incident with any vertex $n\in \mathcal{N}_{T+2}$ is not more than $2$.
Let $x$ be the vector that produces the maximum size $2$-matching in $G[\mathcal{N}_{T+2}]$, i.e., the size of $x$ is 
\begin{align} \sum_{uv\in E[\mathcal{N}_{T+2}]}x(uv)&=\nu_2(G[\mathcal{N}_{T+2}]).
\end{align} 
Multiplying both sides of \eqref{eq:duv} by $x(uv)$, summing up over all $uv\in E[\mathcal{N}_{T+2}]$, and adding $2\times \eqref{eq:dt}$, we have
\begin{align}
\sum_{uv\in E[\mathcal{N}_{T+2}]}(D_u+D_v)x(uv)+2\sum_{t\in[\mathcal{N}_{T+1}]}(D_t)&\geq \sum_{uv\in E[\mathcal{N}_{T+2}]}x(uv)+2|\mathcal{N}_{T+1}|\label{eq:con1deq1}\\
\Rightarrow \sum_{u\in\mathcal{N}_{T+2}}x(\delta(u)\cap E[\mathcal{N}_{T+2}])(D_u)+2\sum_{t\in[\mathcal{N}_{T+1}]}(D_t)&\geq \nu_2(G[\mathcal{N}_{T+2}])+2|\mathcal{N}_{T+1}|\label{eq:con1deq2}\\
\Rightarrow 2\sum_{u\in\mathcal{N}_{T+2}}(D_u)+2\sum_{t\in[\mathcal{N}_{T+1}]}(D_t)&\geq \nu_2(G[\mathcal{N}_{T+2}])+2|\mathcal{N}_{T+1}|\label{eq:con1deq3}\\
\Rightarrow 2\sum_{u\in[N]}(D_u)&\geq\nu_2(G[\mathcal{N}_{T+2}])+2|\mathcal{N}_{T+1}|\label{eq:con1deq4}\\
\Rightarrow (D_1+D_2+\cdots+D_n)&\geq  \frac{\nu_2(G[\mathcal{N}_{T+2}])+2|\mathcal{N}_{T+1}|}{2}\label{eq:con1deq6}
\end{align}
In \eqref{eq:con1deq3} we used the fact that the sum of values assigned by $x$ to all edges in $E[\mathcal{N}_{T+2}]$ that are incident with the vertex $u$ is not more than $2$. 
Combining \eqref{eq:con1deq6} with the result of Theorem \ref{thm:converse},  we obtain the desired converse bound
\begin{align}
C_\infty &\leq \frac{2}{\nu_2(G[\mathcal{N}_{T+2}])+2|\mathcal{N}_{T+1}|}.\label{eq:con1deq7}
\end{align}
Thus, the proof of converse for Theorem \ref{thm:GTPIR} is complete. $\hfill\square$
\subsection{Proof of Achievability for Theorem \ref{thm:GTPIR}}
Let us define $\mathcal{W}_{T+1}$ as the set of messages that are replicated $T+1$ times. Let $U\subset \mathcal{N}_{T+2}$ be a stable set. We will show that it is possible to retrieve $L=2$ desired symbols with a total normalized download,
\begin{align}
D_1+\cdots+D_N&=\frac{|[1:N]\backslash U| + |\mathcal{N}(U)\cup \mathcal{N}_{T+1}|}{2}
\end{align}
The achievable scheme does not use the servers in $U$. Let $\mathcal{W}_U$ denote the set of messages that are stored at any of the servers in $U$. Note that none of these messages is in $\mathcal{W}_{T+1}$ because $U\subset \mathcal{N}_{T+2}$. Also note that no message is replicated more than once in $U$ because $U$ is a stable set. After the servers in $U$ are eliminated, the messages $\mathcal{W}^*=\mathcal{W}_U\cup \mathcal{W}_{T+1}$ are now replicated exactly $(T+1)$ times in the remaining servers. All other messages are replicated $(T+2)$ times. As a thought experiment, suppose we add a genie server that stores $\mathcal{W}^*$. Now we have a storage system where all messages are replicated $(T+2)$ times, so that the scheme presented in the proof of Theorem \ref{thm:ach} can be used to retrieve $L=2$ desired symbols while downloading $|[N]\backslash U| + 1$ symbols, which includes one genie symbol, say $\lambda(\mathcal{W}^*)$. In order to obtain $\lambda(\mathcal{W}^*)$ without a genie, we will use the servers in the set $\mathcal{N}(U)\cup\mathcal{N}_{T+1}$. Note that $\mathcal{N}(U)$ and $\mathcal{N}_{T+1}$ may have some servers in common. More importantly, note that $\mathcal{W}^*$ is replicated $(T+1)$ times within this set. Therefore, we can privately retrieve $\lambda(\mathcal{W}^*)$ by downloading one symbol from each of these servers, with the scheme described in Section \ref{sec:GTPC}. Thus, we have a private and correct scheme that retrieves $L=2$ desired symbols with a total download of $|[N]\backslash U| + |\mathcal{N}(U)\cup \mathcal{N}_{T+1}|$. Next, we note the following identity,
\begin{align}
\underbrace{|[N]\backslash U|}_{t_1} + \underbrace{|\mathcal{N}(U)\cup \mathcal{N}_{T+1}|}_{t_2}&=\underbrace{|\mathcal{N}_{T+2}\backslash U|}_{t_3} +\underbrace{|\mathcal{N}(U)\cap\mathcal{N}_{T+2}|}_{t_4}+\underbrace{2|\mathcal{N}_{T+1}|}_{t_5}\label{eq:identity}
\end{align}
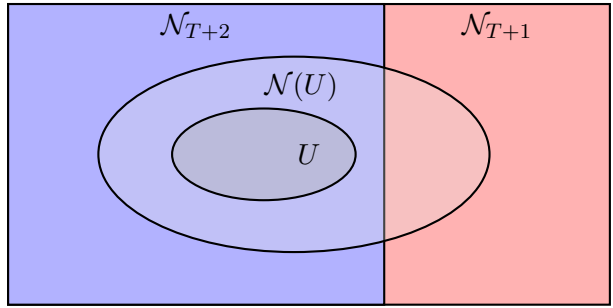
\begin{figure}[h]
\begin{center}
\begin{tikzpicture}
\begin{scope}[xscale=2, yscale=1]
\draw [fill=red!30, thick] (-2, -2) rectangle (2, 2);
\draw [fill=blue!30, thick] (-2, -2) rectangle (0.5, 2);
\draw[fill=black!10, thick, fill opacity=0.3] (-0.1cm,0) circle (1.3cm);
\draw[ thick, fill=black!30, fill opacity=0.4] (-0.3cm,0) circle (0.61cm);
\end{scope}
\node at (0,0) {$U$};
\node at (-0.1,0.9) {$\mathcal{N}(U)$};
\node at (-1.5,1.69) {$\mathcal{N}_{T+2}$};
\node at (+2.5,1.69) {$\mathcal{N}_{T+1}$};
\end{tikzpicture}
\end{center}
\caption{General setting of $U\subset\mathcal{N}_{T+2}$ which may have neighbors $\mathcal{N}(U)$ both in $\mathcal{N}_{T+1}$ and $\mathcal{N}_{T+2}$. Note that $\mathcal{N}(U)$ does not include $U$.}
\end{figure}

\noindent Let us verify that the identity holds  as follows. First consider the servers in $\mathcal{N}_{T+1}$. On the LHS  all these servers are included in $t_1$ as well as $t_2$, i.e., they are counted twice. On the RHS these servers are included only in $t_5$ which is scaled by a factor of $2$, so both sides match. Now consider servers that are in $\mathcal{N}_{T+2}$ and are neighbors of servers in $U$.  On the LHS these servers are included in $t_1$ as well as $t_2$, i.e., they are counted twice. On the RHS, these servers are included in $t_3$ as well as $t_4$, so again they are counted twice and the two sides match. Finally, consider the servers that are in $\mathcal{N}_{T+2}$ but are neither in $U$ nor among the neighbors of the servers in $U$. On the LHS all these servers are included  in $t_1$, while on the RHS they are included in $t_3$. Thus on both sides these servers are included once, and the two sides match. Finally, note that the servers in $U$ are not included in any term on either the LHS or the RHS. Thus, we have verified that \ref{eq:identity} holds.

Now, let us recall that according to \eqref{eq:nu2},
\begin{align}
\nu_2(G[\mathcal{N}_{T+2}])&=\min\{|\mathcal{N}_{T+2}\backslash U| +|\mathcal{N}(U)\cap\mathcal{N}_{T+2}|~\Big|~\mbox{such that } U\subset\mathcal{N}_{T+2}, U \mbox{ is a stable set}\}.
\end{align}
Therefore, minimizing over $U\in\mathcal{N}_{T+2}$, the scheme achieves the normalized download,
\begin{align}
D_1+\cdots+D_N&= \frac{\nu_2(G[\mathcal{N}_{T+2}])}{2}+|\mathcal{N}_{T+1}|,
\end{align}
and therefore we have a lower bound on  capacity,
\begin{align}
C_\infty&\geq \frac{2}{\nu_2(G[\mathcal{N}_{T+2}])+2|\mathcal{N}_{T+1}|}.
\end{align}
Because the achievable scheme works for any number of messages, it is notable that this lower bound holds not only for asymptotic capacity, but also for capacity with arbitrary number of messages $K_m$. This completes the proof of achievability for Theorem \ref{thm:GTPIR}. $\hfill\square$

\section{Conclusion}
The asymptotic capacity of GXSTPIR studied in this work reveals important insights into the structure of optimal schemes for graph-based replicated storage. In particular the special structure inspired by dual GRS codes emerges as a powerful idea for GXSTPIR. Generalizations of the private computation scheme presented in Section \ref{sec:GTPC} represent an interesting problem for future work, especially because such private computation schemes are needed for GXSTPIR, as evident from the achievability proof of Theorem \ref{thm:GTPIR}. Asymptotic capacity for GPIR with arbitrary graph based storage when each message is replicated $4$ times is the next step for the direction initiated by Theorem \ref{thm:GTPIR}. The relationship between GXSTPIR and index coding, through the connecting thread of min-rank problems that arise in both contexts is another promising research avenue. Finally,  the tightness of the converse bound in Theorem \ref{thm:converse} remains an interesting question. Given that the bound is tight in all cases for which the asymptotic capacity is settled so far, it is tempting to conjecture that the converse bound is tight in general. Settling this conjecture is perhaps the most important immediate objective for future work on the asymptotic capacity of GXSTPIR.

\appendix
\section{Lemmas}\label{app:lemmas}
\begin{lemma}\label{lemma:frac} The optimal value of total normalized download, $\min_{\mathcal{D}}(D_1+D_2+\cdots+D_N)$, in Theorem \ref{thm:converse} is equal to the fractional matching number of $\mathcal{G}[\mathcal{V},\mathcal{E}]$.
\end{lemma}
\proof
Let us consider the non-degenerate scenario, $\rho_{\min}>X+T$, because otherwise the asymptotic capacity is zero. According to Theorem \ref{thm:converse}, the optimal value of total normalized download $\min_{\mathcal{D}}(D_1+D_2+\cdots+D_N)$  is expressed as the result of the following linear program.
\begin{align}
D^*=&\min \sum_{n\in[N]}D_n\\
\mbox{such that,}&\\
&\sum_{n: ~n\in e}D_n\geq 1, &\forall e\in\mathcal{E}\\
&D_n\geq 0,&\forall n\in[N]
\end{align}
Since the linear program is bounded and feasible, by the strong duality of linear programming, we have as its dual the following linear program.
\begin{align}
D^*&=\max \sum_{e\in\mathcal{E}}x_e\\
\mbox{such that,}&\\
&\sum_{e:~ e\ni n} x_e\leq 1, &\forall n\in[N]\\
&x_e\geq 0, &\forall e\in\mathcal{E}
\end{align}
Thus, the optimal converse bound $D^*$ is precisely the maximum weight of a fractional $1$-matching in $\mathcal{G}$. Therefore, the converse bound in Theorem \ref{thm:converse} coincides with the achievability bound in Theorem \ref{thm:ach} if and only if $D^*=\frac{N}{\rho_{\min}-X-T}$. This completes the proof of Lemma \ref{lemma:frac}.$\hfill\square$

\begin{lemma}\label{lemma:grs}
For distinct non-zero values $\beta_1, \cdots,\beta_n$
and for $v_1, \cdots, v_n$ defined as
\begin{align}
v_i&\triangleq \left(\prod_{j\in[n]\setminus\{i\}}(\beta_i-\beta_{j})\right)^{-1}, && i\in[n]
\end{align}
the following identity is satisfied,
\begin{align}
\sum_{i\in[n]}v_i\beta_i^{j}&=0, &&\forall j\in\{0,1,\cdots,n-2\}.
\end{align}
\end{lemma}
\proof The proof of Lemma \ref{lemma:grs} follows directly from the properties of dual GRS codes for which we refer the reader to \cite{Macwilliams}. For our purpose let us recall that given two $n$-dimensional vectors
\begin{align}
\mathbf{u}&=[u_1, u_2, \cdots, u_n]\\
\boldsymbol{\beta}&=[\beta_1, \beta_2, \cdots, \beta_n]
\end{align}
where $u_1,u_2, \cdots, u_n$ are non-zero, while $\beta_1, \beta_2, \cdots, \beta_n$ are  non-zero and distinct, the canonical generator matrix for the Generalized Reed-Solomon code $\mbox{GRS}_{k,n}(\mathbf{u},\boldsymbol{\beta})$  is given by
\begin{align}
\left[
\begin{matrix}
u_1& u_2 & \cdots &u_n\\
u_1\beta_1&u_2\beta_2&\cdots&u_n\beta_n\\
\vdots&\vdots&\cdots&\vdots\\
u_1\beta_1^{k-1}&u_2\beta_2^{k-1}&\cdots&u_n\beta_n^{k-1}
\end{matrix}
\right]
\end{align}
The dual code of a GRS code is also a GRS code. Specifically, the dual for $\mbox{GRS}_{k,n}(\mathbf{u},\boldsymbol{\beta})$ is  $\mbox{GRS}_{n-k,n}(\mathbf{v},\boldsymbol{\beta})$
where $\mathbf{v}=[v_1, v_2, \cdots, v_n]$ and $v_i=\left(u_i\prod_{j\in[n]\setminus\{i\}}(\beta_i-\beta_j)\right)^{-1}$. For the purpose of Lemma \ref{lemma:grs} let us set  $u_1=u_2=\cdots=u_n=1$.  Since the dual of  a code $C$ is a code $C^\perp$ that spans the null space of $C$, we have
\begin{align}
\left[
\begin{matrix}
v_1& v_2 & \cdots &v_n\\
v_1\beta_1&v_2\beta_2&\cdots&v_n\beta_n\\
\vdots&\vdots&\cdots&\vdots\\
v_1\beta_1^{k-1}&v_2\beta_2^{k-1}&\cdots&v_n\beta_n^{k-1}
\end{matrix}
\right]\left[
\begin{matrix}
1&\beta_1&\cdots&\beta_1^{n-k-1}\\
1&\beta_2&\cdots&\beta_2^{n-k-1}\\
\vdots&\vdots&\cdots&\vdots\\
1&\beta_n&\cdots&\beta_n^{n-k-1}
\end{matrix}
\right]&={\bf 0}
\end{align}
which implies that
\begin{align}
\sum_{i\in[n]}v_i\beta_i^j&=0
\end{align}
for $j\in\{0,1,\cdots, n-2\}$. This completes the proof of Lemma \ref{lemma:grs}.$\hfill\square$

\bigskip
\begin{lemma}\label{lemma:indSWq}
For all $m\in[M], k\in[K_m]$, $\mathcal{X}\subset\mathcal{R}_m$, $|\mathcal{X}|\leq X$, 
\begin{align}
I\left(S_{[N]\backslash\mathcal{R}_m},S_{\mathcal{X}}; W_{m,k}, Q_{[N]}^{[m,k]}\right)&=0.
\end{align}
\end{lemma}
\proof
\begin{align}
&I(S_{[N]\backslash\mathcal{R}_m},S_{\mathcal{X}}; W_{m,k}, Q_{[N]}^{[m,k]})\\
&=I(W_{m,k};S_{[N]\backslash\mathcal{R}_m},S_{\mathcal{X}})+I(Q_{[N]}^{[m,k]};S_{[N]\backslash\mathcal{R}_m},S_{\mathcal{X}}\mid W_{m,k})\label{eq:conneq1}\\
&\leq I(W_{m,k};S_{[N]\backslash\mathcal{R}_m},S_{\mathcal{X}})+I(Q_{[N]}^{[m,k]};S_{[N]\backslash\mathcal{R}_m},S_{\mathcal{X}},W_{m,k})\label{eq:conneq2}\\
&=I(W_{m,k};S_{[N]\backslash\mathcal{R}_m},S_{\mathcal{X}})\label{eq:conneq3}\\
&\leq I(W_{m,k};\overline{\mathcal{W}}', \overline{W}_{m,k}^{(\mathcal{X})})\label{eq:conneq4}\\
&=I(W_{m,k};\overline{W}_{m,k}^{(\mathcal{X})})+I(W_{m,k};\overline{\mathcal{W}}'|\overline{W}_{m,k}^{(\mathcal{X})})\label{eq:conneq5}\\
&=I(W_{m,k};\overline{\mathcal{W}}'|\overline{W}_{m,k}^{(\mathcal{X})})\label{eq:conneq6}\\
&\leq I(W_{m,k},\overline{W}_{m,k}^{(\mathcal{X})};\overline{\mathcal{W}}')\label{eq:conneq7}\\
&\leq I(\overline{W}_{m,k};\overline{\mathcal{W}}')\label{eq:conneq8}\\
&=0.\label{eq:conneq9}
\end{align}
where $\overline{\mathcal{W}}'=(\overline{W}_{m',k'}, \forall m'\in[M], k'\in[K_m], (m',k')\neq (m,k))$, and $\overline{W}_{m,k}^{(\mathcal{X})}=(\overline{W}_{m,k}^{(n)}, n\in\mathcal{X})$. Steps  of the proof are justified as follows. \eqref{eq:conneq1} and \eqref{eq:conneq2} follow from the chain rule and the non-negativity of mutual information. 
\eqref{eq:conneq3} follows from \eqref{eq:indp}, while \eqref{eq:conneq4}, follows from the definition of replicated storage in \eqref{eq:stor}. \eqref{eq:conneq5} is the chain rule of mutual information, while \eqref{eq:conneq6} follows from the security constraint in \eqref{eq:secur}. \eqref{eq:conneq7} follows from chain rule and the non-negativity of mutual information. In \eqref{eq:conneq8} we used the fact that $(W_{m,k},\overline{W}_{m,k}^{(\mathcal{X})})$ is function of $\overline{W}_{m,k}$, and the last step follows from \eqref{shareindp}. 
This completes the proof of Lemma \ref{lemma:indSWq}.$\hfill\square$

\begin{lemma}\label{lemma:NtoT}
For all $m\in[M], k\in[K_m]$, $\mathcal{X},\mathcal{T}\subset\mathcal{R}_m$, 
\begin{align}
I(W_{m,k}; A_{\mathcal{T}}^{[m,k]}\mid S_{\mathcal{X}}, S_{[N]\backslash\mathcal{R}_m}, Q_{[N]}^{[m,k]})&\leq I(W_{m,k}; A_{\mathcal{T}}^{[m,k]}\mid S_{\mathcal{X}}, S_{[N]\backslash\mathcal{R}_m}, Q_{\mathcal{T}}^{[m,k]}).
\end{align}
\end{lemma}
\proof
\begin{align}
&I(W_{m,k}; A_{\mathcal{T}}^{[m,k]}\mid S_{\mathcal{X}}, S_{[N]\backslash\mathcal{R}_m}, Q_{[N]}^{[m,k]})\notag\\
&=H(A_{\mathcal{T}}^{[m,k]}\mid S_{\mathcal{X}}, S_{[N]\backslash\mathcal{R}_m}, Q_{[N]}^{[m,k]})-H(A_{\mathcal{T}}^{[m,k]}\mid W_{m,k}, S_{\mathcal{X}}, S_{[N]\backslash\mathcal{R}_m}, Q_{[N]}^{[m,k]})\label{eq:con1aeq1}\\
&\leq H(A_{\mathcal{T}}^{[m,k]}\mid S_{\mathcal{X}}, S_{[N]\backslash\mathcal{R}_m}, Q_{\mathcal{T}}^{[m,k]})-H(A_{\mathcal{T}}^{[m,k]}\mid W_{m,k}, S_{\mathcal{X}}, S_{[N]\backslash\mathcal{R}_m}, Q_{[N]}^{[m,k]})\label{eq:con1aeq2}\\
&=H(A_{\mathcal{T}}^{[m,k]}\mid S_{\mathcal{X}}, S_{[N]\backslash\mathcal{R}_m}, Q_{\mathcal{T}}^{[m,k]})-H(A_{\mathcal{T}}^{[m,k]}\mid W_{m,k}, S_{\mathcal{X}}, S_{[N]\backslash\mathcal{R}_m}, Q_{\mathcal{T}}^{[m,k]})\notag\\
&\hspace{1cm}+H(A_{\mathcal{T}}^{[m,k]}\mid W_{m,k}, S_{\mathcal{X}}, S_{[N]\backslash\mathcal{R}_m}, Q_{\mathcal{T}}^{[m,k]})-H(A_{\mathcal{T}}^{[m,k]}\mid W_{m,k}, S_{\mathcal{X}}, S_{[N]\backslash\mathcal{R}_m}, Q_{[N]}^{[m,k]})\label{eq:con1aeq3}\\
&=I(W_{m,k}; A_{\mathcal{T}}^{[m,k]}\mid S_{\mathcal{X}}, S_{[N]\backslash\mathcal{R}_m}, Q_{\mathcal{T}}^{[m,k]})+I(A_{\mathcal{T}}^{[m,k]}; Q_{[N]}^{[m,k]}\mid W_{m,k}, S_{\mathcal{X}}, S_{[N]\backslash\mathcal{R}_m}, Q_{\mathcal{T}}^{[m,k]})\label{eq:con1aeq4}\\
&\leq I(W_{m,k}; A_{\mathcal{T}}^{[m,k]}\mid S_{\mathcal{X}}, S_{[N]\backslash\mathcal{R}_m}, Q_{\mathcal{T}}^{[m,k]})+I(A_{\mathcal{T}}^{[m,k]}, W_{m,k}, S_{\mathcal{X}}, S_{[N]\backslash\mathcal{R}_m}; Q_{[N]}^{[m,k]}\mid Q_{\mathcal{T}}^{[m,k]})\label{eq:con1aeq5}\\
&\leq I(W_{m,k}; A_{\mathcal{T}}^{[m,k]}\mid S_{\mathcal{X}}, S_{[N]\backslash\mathcal{R}_m}, Q_{\mathcal{T}}^{[m,k]})+I(A_{\mathcal{T}}^{[m,k]}, S_{[N]}; Q_{[N]}^{[m,k]}\mid Q_{\mathcal{T}}^{[m,k]})\label{eq:con1aeq6}\\
&=I(W_{m,k}; A_{\mathcal{T}}^{[m,k]}\mid S_{\mathcal{X}}, S_{[N]\backslash\mathcal{R}_m}, Q_{\mathcal{T}}^{[m,k]})+I(S_{[N]}; Q_{[N]}^{[m,k]}\mid Q_{\mathcal{T}}^{[m,k]})+I(A_{\mathcal{T}}^{[m,k]}; Q_{[N]}^{[m,k]}\mid S_{[N]}, Q_{\mathcal{T}}^{[m,k]})\label{eq:con1aeq7}\\
&=I(W_{m,k}; A_{\mathcal{T}}^{[m,k]}\mid S_{\mathcal{X}}, S_{[N]\backslash\mathcal{R}_m}, Q_{\mathcal{T}}^{[m,k]})\label{eq:con1aeq8}
\end{align}
\eqref{eq:con1aeq1} follows from the definition of mutual information, \eqref{eq:con1aeq2} because dropping conditioning cannot reduce entropy, \eqref{eq:con1aeq3} adds and subtracts the same term so nothing changes, \eqref{eq:con1aeq4} uses the definition of mutual information, \eqref{eq:con1aeq5} uses the chain rule of mutual information and the fact that mutual information is always non-negative,  \eqref{eq:con1aeq6} uses the fact that $\left(W_{m,k}, S_{\mathcal{X}}, S_{[N]\backslash\mathcal{R}_m}\right)$ is a function of $S_{[N]}$ according to  \eqref{eq:wfromshare} and \eqref{eq:stor}, and \eqref{eq:con1aeq7} uses chain rule of mutual information. For \eqref{eq:con1aeq8} we use the fact that $S_{[N]}$ is independent of $Q_{[N]}^{[m,k]}$ according to \eqref{eq:indp}, and $A_{\mathcal{T}}^{[m,k]}$ is fully determined by $S_{[N]}, Q_{\mathcal{T}}^{[m,k]}$ according to \eqref{eq:ansfunc}. This completes the proof of Lemma \ref{lemma:NtoT}.$\hfill\square$

\begin{lemma}\label{lemma:condpriv}
For any $m\in[M]$, $\mathcal{T}\subset\mathcal{R}_m$, $|\mathcal{T}|\leq T$,
\begin{align}
I(Q_{\mathcal{T}}^{[m,\kappa]}, A_{\mathcal{T}}^{[m,\kappa]}, S_{[N]}; \kappa)&=0
\end{align}
\end{lemma}
\proof
\begin{align}
I(Q_{\mathcal{T}}^{[m,\kappa]}, A_{\mathcal{T}}^{[m,\kappa]}, S_{[N]}; \kappa)&=I(Q_{\mathcal{T}}^{[m,\kappa]}; \kappa)+I(S_{[N]}; \kappa\mid Q_{\mathcal{T}}^{[m,\kappa]})+I(A_{\mathcal{T}}^{[m,\kappa]}; \kappa\mid S_{[N]},Q_{\mathcal{T}}^{[m,\kappa]})\label{eq:con1beq1}\\
&=I(Q_{\mathcal{T}}^{[m,\kappa]}; \kappa)+I(S_{[N]}; \kappa\mid Q_{\mathcal{T}}^{[m,\kappa]})\label{eq:con1beq2}\\
&\leq I(Q_{\mathcal{T}}^{[m,\kappa]}; \kappa)+I(S_{[N]}; \kappa, Q_{\mathcal{T}}^{[m,\kappa]})\label{eq:con1beq3} \\
&=0\label{eq:con1beq4}
\end{align}
\eqref{eq:con1beq1} is the chain rule of mutual information, \eqref{eq:con1beq2} follows because $A_{\mathcal{T}}^{[\mu,\kappa]}$ is fully determined by $S_{[N]}, Q_{\mathcal{T}}^{[\mu,\kappa]}$ according to \eqref{eq:ansfunc}. The next step, \eqref{eq:con1beq3} follows because of the chain rule of mutual information and the non-negativity of mutual information, and \eqref{eq:con1beq4} follows from \eqref{eq:indp},\eqref{eq:tpriv}. This completes the proof of Lemma \ref{lemma:condpriv}.$\hfill\square$

\begin{lemma}\label{lemma:condind}
For any $m\in[M], k\in[K_m]$ and  subsets $\mathcal{X}, \mathcal{T}\subset\mathcal{R}_m$ such that $|\mathcal{X}|\leq X$, 
\begin{align}
I\left(\mathcal{W}_{m,\mathcal{K}}~;~\mathcal{W}_{m,\mathcal{K}'}\mid S_{\mathcal{X}}, S_{[N]\setminus\mathcal{R}_m},Q_{\mathcal{T}}^{[m,k]}\right)&=0
\end{align}
where $\mathcal{K}\subset [K_m]$, $\mathcal{K}'=[K_m]\setminus\mathcal{K}$, $\mathcal{W}_{m,\mathcal{K}}=(W_{m,k}, k\in\mathcal{K})$ and $\mathcal{W}_{m,\mathcal{K}'}=(W_{m,k}, k\in\mathcal{K}')$.
\end{lemma}
\proof Let us define $\overline{\mathcal{W}}_{\mathcal{M}'}=(\overline{W}_{m',k},\forall m'\in[M], k\in [K_{m'}], m'\neq m)$. $\overline{\mathcal{W}}_{m,\mathcal{K}}=(\overline{W}_{m,k}, k\in\mathcal{K})$. $\overline{\mathcal{W}}_{m,\mathcal{K}'}=(\overline{W}_{m,k}, k\in\mathcal{K}')$. $\overline{\mathcal{W}}_{m,\mathcal{K}}^{(\mathcal{X})}=(\overline{\mathcal{W}}_{m,k}^{(n)}, n\in\mathcal{X},k\in\mathcal{K})$. $\overline{\mathcal{W}}_{m,\mathcal{K}'}^{(\mathcal{X})}=(\overline{\mathcal{W}}_{m,k}^{(n)}, n\in\mathcal{X},k\in\mathcal{K}')$.
\begin{align}
&I(\mathcal{W}_{m,\mathcal{K}};\mathcal{W}_{m,\mathcal{K}'}\mid S_{\mathcal{X}},S_{[N]\setminus\mathcal{R}_m},Q_{\mathcal{T}}^{[m,k']})\\
&\leq I(\mathcal{W}_{m,\mathcal{K}};\mathcal{W}_{m,\mathcal{K}'},S_{\mathcal{X}},S_{[N]\setminus\mathcal{R}_m},Q_{\mathcal{T}}^{[m,k']})\label{eq:conineq1}\\
&= I(\mathcal{W}_{m,\mathcal{K}};\mathcal{W}_{m,\mathcal{K}'},S_{\mathcal{X}},S_{[N]\setminus\mathcal{R}_m})+I(\mathcal{W}_{m,\mathcal{K}};Q_{\mathcal{T}}^{[m,k']}\mid \mathcal{W}_{m,\mathcal{K}'},S_{\mathcal{X}},S_{[N]\setminus\mathcal{R}_m})\label{eq:conineq2}\\
&\leq I(\mathcal{W}_{m,\mathcal{K}};\mathcal{W}_{m,\mathcal{K}'},S_{\mathcal{X}},S_{[N]\setminus\mathcal{R}_m})+I(Q_{\mathcal{T}}^{[m,k']};\mathcal{W}_{m,\mathcal{K}},\mathcal{W}_{m,\mathcal{K}'},S_{\mathcal{X}},S_{[N]\setminus\mathcal{R}_m})\label{eq:conineq3}\\
&=I(\mathcal{W}_{m,\mathcal{K}};\mathcal{W}_{m,\mathcal{K}'},S_{\mathcal{X}},S_{[N]\setminus\mathcal{R}_m})\label{eq:conineq4}\\
&\leq I(\mathcal{W}_{m,\mathcal{K}};\mathcal{W}_{m,\mathcal{K}'},\overline{\mathcal{W}}_{\mathcal{M}'},\overline{\mathcal{W}}_{m,\mathcal{K}}^{(\mathcal{X})},\overline{\mathcal{W}}_{m,\mathcal{K}'}^{(\mathcal{X})})\label{eq:conineq5}\\
&\leq I(\mathcal{W}_{m,\mathcal{K}};\overline{\mathcal{W}}_{m,\mathcal{K}'},\overline{\mathcal{W}}_{\mathcal{M}'},\overline{\mathcal{W}}_{m,\mathcal{K}}^{(\mathcal{X})})\label{eq:conineq6}\\
&=I(\mathcal{W}_{m,\mathcal{K}};\overline{\mathcal{W}}_{m,\mathcal{K}}^{(\mathcal{X})})+I(\mathcal{W}_{m,\mathcal{K}};\overline{\mathcal{W}}_{m,\mathcal{K}'},\overline{\mathcal{W}}_{\mathcal{M}'}\mid \overline{\mathcal{W}}_{m,\mathcal{K}}^{(\mathcal{X})})\label{eq:conineq7}\\
&=I(\mathcal{W}_{m,\mathcal{K}};\overline{\mathcal{W}}_{m,\mathcal{K}'},\overline{\mathcal{W}}_{\mathcal{M}'}\mid \overline{\mathcal{W}}_{m,\mathcal{K}}^{(\mathcal{X})})\label{eq:conineq8}\\
&\leq I(\mathcal{W}_{m,\mathcal{K}},\overline{\mathcal{W}}_{m,\mathcal{K}}^{(\mathcal{X})};\overline{\mathcal{W}}_{m,\mathcal{K}'},\overline{\mathcal{W}}_{\mathcal{M}'})\label{eq:conineq9}\\
&\leq I(\overline{\mathcal{W}}_{m,\mathcal{K}};\overline{\mathcal{W}}_{m,\mathcal{K}'},\overline{\mathcal{W}}_{\mathcal{M}'})\label{eq:conineq10}\\
&=0.\label{eq:conineq11}
\end{align}
 \eqref{eq:conineq1}, \eqref{eq:conineq2}, \eqref{eq:conineq3} follows from the chain rule and the non-negativity of mutual information. \eqref{eq:conineq4} holds because of \eqref{eq:indp}, while in \eqref{eq:conineq5}, we used the definition of the storage as in \eqref{eq:stor}. \eqref{eq:conineq6} follows because $\left(\mathcal{W}_{m,\mathcal{K}'},\overline{\mathcal{W}}_{m,\mathcal{K}'}^{(\mathcal{X})}\right)$ is function of $\overline{\mathcal{W}}_{m,\mathcal{K}'}$. \eqref{eq:conineq7} is again the chain rule of mutual information, and \eqref{eq:conineq8} follows from the $X$-security constraint as in \eqref{eq:secur}. \eqref{eq:conineq9} follows from the chain rule and the non-negativity of mutual information, while in \eqref{eq:conineq10}, we used the fact that $\left(\mathcal{W}_{m,\mathcal{K}},\overline{\mathcal{W}}_{m,\mathcal{K}}^{(\mathcal{X})}\right)$ is function of $\overline{\mathcal{W}}_{m,\mathcal{K}}$. The last step holds because of \eqref{shareindp}. This completes the proof of Lemma \ref{lemma:condind}. $\hfill\square$

\medskip

\bibliographystyle{IEEEtran}
\bibliography{Thesis}

\end{document}